\documentclass[12pt]{article}
\usepackage{amsmath,amssymb}
\usepackage{graphicx}
\newcommand{\eqdef}{\stackrel{\text{def}}{=}}

\newcommand{\n}{\nonumber\\}
\newcommand{\bm}{\boldsymbol}
\newcommand{\ignore}[1]{}
\numberwithin{equation}{section}
\newcommand{\Romannumeral}[1]{\uppercase\expandafter{\romannumeral#1}}
\newcommand{\I}{\text{\Romannumeral{1}}}
\newcommand{\II}{\text{\Romannumeral{2}}}

\newcommand{\cB}{\mathcal{B}}
\newcommand{\cD}{\mathcal{D}}
\newcommand{\cE}{\mathcal{E}}
\newcommand{\tcE}{\tilde{\mathcal{E}}}
\newcommand{\cF}{\mathcal{F}}
\newcommand{\cH}{\mathcal{H}}
\newcommand{\wtcH}{\widetilde{\mathcal{H}}}
\newcommand{\cP}{\mathcal{P}}
\newcommand{\cX}{\mathcal{X}}

\allowdisplaybreaks[4]

\begin{document}

\baselineskip=20pt

\newcommand{\preprint}{
\vspace*{-20mm}
   \begin{flushright}\normalsize \sf
    DPSU-25-1\\
  \end{flushright}}
\newcommand{\Title}[1]{{\baselineskip=26pt
  \begin{center} \Large \bf #1 \\ \ \\ \end{center}}}
\newcommand{\Author}{\begin{center}
  \large \bf Satoru Odake \end{center}}
\newcommand{\Address}{\begin{center}
     Faculty of Science, Shinshu University,
     Matsumoto 390-8621, Japan
   \end{center}}
\newcommand{\Accepted}[1]{\begin{center}
  {\large \sf #1}\\ \vspace{1mm}{\small \sf Accepted for Publication}
  \end{center}}

\preprint
\thispagestyle{empty}

\Title{Multi-indexed Orthogonal Polynomials of a Discrete Variable and
Exactly Solvable Birth and Death Processes}

\Author

\Address
\vspace{1cm}

\begin{abstract}
We present the case-(1) multi-indexed orthogonal polynomials of a discrete
variable for 8 types ((dual)($q$-)Hahn, three kinds of $q$-Krawtchouk and
$q$-Meixner).
Based on them and the case-(1) multi-indexed orthogonal polynomials of 
Racah, $q$-Racah, Meixner, little $q$-Jacobi and little $q$-Laguerre types,
exactly solvable continuous time birth and death processes are obtained.
Their discrete time versions (Markov chains) are also obtained for finite
types.
\end{abstract}

\section{Introduction}
\label{sec:intro}

The exceptional and multi-indexed orthogonal polynomials are new type of
orthogonal polynomials \cite{gkm08,os16,os25,os26,os27,ggm13,d14,d17}.
They form a complete set of orthogonal basis in spite of the missing degrees,
by which the restrictions of Bochner's theorem \cite{ismail} are avoided.
We distinguish the following two cases; the set of missing degrees is
case-(1): $\{0,1,\ldots,\ell-1\}$, where $\ell$ is a positive integer, and
case-(2): otherwise.
They are constructed based on the polynomials in the Askey-scheme of
hypergeometric orthogonal polynomials \cite{kls}, which satisfy second order
differential or difference equations. In the study of such orthogonal
polynomials, we use the quantum mechanical formulation \cite{os24}.
In this paper we consider orthogonal polynomials of a discrete variable
\cite{ismail,kls,nsu}.
To study them, we use the discrete quantum mechanics with real shifts (rdQM)
\cite{os12,os34}.
The multi-indexed orthogonal polynomials of a discrete variable are studied in
\cite{os22,os23,os26,d14,d17,mtv22b,os40,addrdQM}.
The Krein-Adler type multi-indexed orthogonal polynomials \cite{os22,addrdQM}
are the case-(2) polynomials.
The multi-indexed orthogonal polynomials studied in \cite{mtv22b,os40,addrdQM}
have added degrees.
The case-(1) multi-indexed polynomials are constructed for Racah and $q$-Racah
types \cite{os26} and for Meixner, little $q$-Jacobi and little $q$-Laguerre
types \cite{os35,lqJII}.
The deformed quantum systems described by these case-(1) polynomials have
shape invariant property.

The first purpose of this paper is to expand the list of the case-(1)
multi-indexed orthogonal polynomials of a discrete variable.
By the same methods in \cite{os26,os35} (with some modification), we obtain
the multi-indexed orthogonal polynomials of 8 types ((dual)($q$-)Hahn, three
kinds of $q$-Krawtchouk and $q$-Meixner).

The second purpose of this paper is an application of the multi-indexed
orthogonal polynomials.
Orthogonal polynomials have various application \cite{ismail,nsu}, and it is
an important problem to clarify whether such applications can be extended to
the multi-indexed orthogonal polynomials.
In this paper, we consider the birth and death (BD) processes
\cite{km57a,km57b,ismail,s09,s21,mtv22a}.
For each orthogonal polynomials of a discrete variable in the Askey-scheme,
$\check{P}_n(x)$ ($x\in\cX$, \eqref{cXdef}), the exactly solvable BD processes
(with continuous time) are nicely obtained by Sasaki \cite{s09}.
The polynomials $\check{P}_n(x)$ satisfy the difference equation
\begin{equation*}
  \bigl(B(x)+D(x)\bigr)\check{P}_n(x)-B(x)\check{P}_n(x+1)-D(x)\check{P}_n(x-1)
  =\cE_n\check{P}_n(x).
\end{equation*}
The sum of the coefficients of $\check{P}_n(x)$, $\check{P}_n(x+1)$ and
$\check{P}_n(x-1)$ in the left hand side is
\begin{equation*}
  \bigl(B(x)+D(x)\bigr)-B(x)-D(x)=0.
\end{equation*}
This relation and the boundary conditions
ensure the conservation of probability of the BD process.
Moreover, from these continuous time BD processes, the discrete time BD
processes (Markov chain) are also obtained by Sasaki \cite{s21}.
The case-(1) multi-indexed orthogonal polynomials
$\check{P}_{\cD,n}(x)=\check{P}_{\cD,n}(x;\bm{\lambda})$
satisfy the difference equation
\begin{align*}
  &\quad B(x;\bm{\lambda}+M\tilde{\bm{\delta}})
  \frac{\check{\Xi}_{\cD}(x;\bm{\lambda})}
  {\check{\Xi}_{\cD}(x+1;\bm{\lambda})}
  \Bigl(\frac{\check{\Xi}_{\cD}(x+1;\bm{\lambda}+\bm{\delta})}
  {\check{\Xi}_{\cD}(x;\bm{\lambda}+\bm{\delta})}
  \check{P}_{\cD,n}(x;\bm{\lambda})-\check{P}_{\cD,n}(x+1;\bm{\lambda})\Bigr)\n
  &+D(x;\bm{\lambda}+M\tilde{\bm{\delta}})
  \frac{\check{\Xi}_{\cD}(x+1;\bm{\lambda})}
  {\check{\Xi}_{\cD}(x;\bm{\lambda})}
  \Bigl(\frac{\check{\Xi}_{\cD}(x-1;\bm{\lambda}+\bm{\delta})}
  {\check{\Xi}_{\cD}(x;\bm{\lambda}+\bm{\delta})}
  \check{P}_{\cD,n}(x;\bm{\lambda})-\check{P}_{\cD,n}(x-1;\bm{\lambda})\Bigr)\n
  &=\cE_n(\bm{\lambda})\check{P}_{\cD,n}(x;\bm{\lambda}).
\end{align*}
The sum of the coefficients of $\check{P}_{\cD,n}(x;\bm{\lambda})$,
$\check{P}_{\cD,n}(x+1;\bm{\lambda})$ and $\check{P}_{\cD,n}(x-1;\bm{\lambda})$
in the left hand side,
\begin{align*}
  &\Bigl(B(x;\bm{\lambda}+M\tilde{\bm{\delta}})
  \frac{\check{\Xi}_{\cD}(x;\bm{\lambda})}
  {\check{\Xi}_{\cD}(x+1;\bm{\lambda})}
  \frac{\check{\Xi}_{\cD}(x+1;\bm{\lambda}+\bm{\delta})}
  {\check{\Xi}_{\cD}(x;\bm{\lambda}+\bm{\delta})}\n
  &\quad+D(x;\bm{\lambda}+M\tilde{\bm{\delta}})
  \frac{\check{\Xi}_{\cD}(x+1;\bm{\lambda})}
  {\check{\Xi}_{\cD}(x;\bm{\lambda})}
  \frac{\check{\Xi}_{\cD}(x-1;\bm{\lambda}+\bm{\delta})}
  {\check{\Xi}_{\cD}(x;\bm{\lambda}+\bm{\delta})}\Bigr)\n
  &-B(x;\bm{\lambda}+M\tilde{\bm{\delta}})
  \frac{\check{\Xi}_{\cD}(x;\bm{\lambda})}
  {\check{\Xi}_{\cD}(x+1;\bm{\lambda})}
  -D(x;\bm{\lambda}+M\tilde{\bm{\delta}})
  \frac{\check{\Xi}_{\cD}(x+1;\bm{\lambda})}
  {\check{\Xi}_{\cD}(x;\bm{\lambda})},
\end{align*}
does not vanish (is not a constant) in general.
We found that this sum becomes a constant for ($q$-)Racah types with special
parameters $\bm{\lambda}$ and index set $\mathcal{D}$,
but the boundary conditions are not satisfied.
So we have thought that the BD processes associated with the multi-indexed
orthogonal polynomials are impossible.
However, this difficulty can be overcome by considering the ratio of the
polynomials instead of the polynomials.
We have obtained exactly solvable BD processes associated with
the multi-indexed orthogonal polynomials.
Their discrete time versions are also obtained.

This paper is organized as follows.
In section \ref{sec:miop}, the case-(1) multi-indexed orthogonal polynomials
of a discrete variable are recapitulated and some similarity transformed
Hamiltonians are presented. The case-(1) multi-indexed orthogonal polynomials
of 8 types (Hahn etc.) are new results.
In section \ref{sec:BD}, exactly solvable BD processes associated with the
multi-indexed orthogonal polynomials are obtained for both continuous and
discrete times. The repeated discrete time BD processes and its continuous
time version are also obtained.
Section \ref{sec:summary} is for a summary and comments.
In Appendix \ref{app:data} the basic data of the case-(1) multi-indexed
orthogonal polynomials are presented.

\section{Multi-indexed Orthogonal Polynomials}
\label{sec:miop}

In this section we recapitulate the case-(1) multi-indexed orthogonal
polynomials of a discrete variable and present their concrete forms for
the known 5 types and new 8 types.
We also present some similarity transformed Hamiltonians.

The case-(1) multi-indexed orthogonal polynomials of a discrete variable
are constructed by using quantum mechanical formulation, rdQM on a lattice
$\cX$,
\begin{equation}
  \cX\eqdef
  \begin{cases}
  \{0,1,\ldots,N\}&:\text{finite system}\\
  \ \mathbb{Z}_{\geq0}&:\text{semi-infinite system}
  \end{cases}.
  \label{cXdef}
\end{equation}
For finite rdQM systems, the Racah (R) and $q$-Racah ($q$R) types are obtained
in \cite{os26}, and for semi-infinite rdQM systems, the Meixner (M), little
$q$-Jacobi (l$q$J) and little $q$-Laguerre (l$q$L) types are obtained in
\cite{os35}. 
By the same methods in \cite{os26,os35} (with some modification), the case-(1)
multi-indexed orthogonal polynomials of the
Hahn (H),
dual Hahn (dH),
dual quantum $q$-Krawtchouk (dq$q$K),
$q$-Hahn ($q$H),
quantum $q$-Krawtchouk (q$q$K),
affine $q$-Krawtchouk (a$q$K),
dual $q$-Hahn (d$q$H)
and $q$-Meixner ($q$M)
types are concretely constructed in this paper.
We present their data in Appendix \ref{app:data}.
We consider these 13 types of multi-indexed orthogonal polynomials.
Although the type $\II$ multi-indexed little $q$-Jacobi and $q$-Laguerre
polynomials constructed in \cite{lqJII} are the case-(1) polynomials,
we do not treat them here, because their some expressions are slightly
different.

Our notation and the common quantities of the case-(1) multi-indexed
polynomials are summarized in \S\,\ref{sec:common}.
We have the following properties:
\begin{align}
  &\check{\Xi}_{\cD}(0;\bm{\lambda})=1,\quad
  \check{P}_{\cD,n}(0;\bm{\lambda})=1,\quad
  \psi_{\cD}(0;\bm{\lambda})=1,\quad
  \phi_{\cD\,n}(0;\bm{\lambda})=1,
  \label{XiD(0)PDn(0)}\\
  &\Xi_{\cD}(\eta;\bm{\lambda})
  \text{ : a polynomial of degree $\ell_{\cD}$ in $\eta$},
  \label{XiDdeg}\\
  &P_{\cD,n}(\eta;\bm{\lambda})
  \text{ : a polynomial of degree $\ell_{\cD}+n$ in $\eta$},
  \label{PDndeg}\\
  &\check{P}_{\cD,0}(x;\bm{\lambda})
  =\check{\Xi}_{\cD}(x;\bm{\lambda}+\bm{\delta}).
  \label{PD0=XiD}
\end{align}
The property \eqref{PDndeg} means that the set of missing degrees is
$\{0,1,\ldots,\ell_{\mathcal{D}}-1\}$, namely the case-(1) polynomials.
The basic data for each polynomial are given in \S\,\ref{sec:eachpoly},
and 8 types (H, $q$H, etc.) are new results.

The Schr\"{o}dinger equation of rdQM is a matrix eigenvalue problem.
The Hamiltonian $\cH_{\cD}$ is a real symmetric matrix,
\begin{align}
  \cH_{\cD}(\bm{\lambda})
  &=\bigl(\cH_{\cD}(\bm{\lambda})_{x,y}\bigr)_{x,y\in\cX}\,,\\
  \cH_{\cD}(\bm{\lambda})_{x,y}&\eqdef
  \bigl(B_{\cD}(x;\bm{\lambda})+D_{\cD}(x;\bm{\lambda})\bigr)\delta_{x,y}\n
  &\quad
  -\sqrt{B_{\cD}(x;\bm{\lambda})D_{\cD}(x+1;\bm{\lambda})}\,\delta_{x+1,y}
  -\sqrt{B_{\cD}(x-1;\bm{\lambda})D_{\cD}(x;\bm{\lambda})}\,\delta_{x-1,y}\,.
\end{align}
Here the potential functions $B_{\cD}(x)$ \eqref{BDdef} and
$D_{\cD}(x)$ \eqref{DDdef} are
positive except for one boundary,
\begin{align}
  &\text{finite case :}
  \ \ B_{\cD}(x;\bm{\lambda})>0\ \ (x\in\{0,1,\ldots,N-1\}),
  \ \ B_{\cD}(N;\bm{\lambda})=0,\n
  &\phantom{\text{finite case :}\ \ }
  D_{\cD}(0;\bm{\lambda})=0,
  \ \ D_{\cD}(x;\bm{\lambda})>0\ \ (x\in\{1,2,\ldots,N\}),
  \label{BD,DD>0}\\
  &\text{semi-infinite case :}
  \ \ B_{\cD}(x;\bm{\lambda})>0\ \ (x\in\mathbb{Z}_{\geq0}),\n
  &\phantom{\text{semi-infinite case :}\ \ }
  D_{\cD}(0;\bm{\lambda})=0,
  \ \ D_{\cD}(x;\bm{\lambda})>0\ \ (x\in\mathbb{Z}_{\geq1}),
  \label{BD,DD>0inf}
\end{align}
for appropriate parameter ranges given in \S\,\ref{sec:eachpoly}.
For these parameter ranges, the denominator polynomial
$\check{\Xi}_{\cD}(x)$ \eqref{XiDdef} is positive on $\cX$,
$\check{\Xi}_{\cD}(x;\bm{\lambda})>0$ ($x\in\cX$ and $x=N+1$ for finite
systems),
and the multi-indexed orthogonal polynomial $P_{\cD,n}(\eta;\bm{\lambda})$
\eqref{PDndef} has $n$ zeros in the physical region
$0\leq\eta\leq\eta(N;\bm{\lambda}+M\tilde{\bm{\delta}})$
($\Leftrightarrow x\in[0,N]$) for finite cases
($0\leq\eta$ ($\Leftrightarrow x\in\mathbb{R}_{\geq 0}$) for M and $q$M cases,
$0\leq\eta<1$ ($\Leftrightarrow x\in\mathbb{R}_{\geq 0}$) for l$q$J and l$q$L
cases),
and $\ell_{\cD}$ zeros in the unphysical region
$\eta\in\mathbb{C}\backslash[0,\eta(N;\bm{\lambda}+M\tilde{\bm{\delta}})]$
for finite cases
($\eta\in\mathbb{C}\backslash\mathbb{R}_{\geq0}$ for M and $q$M cases,
$\eta\in\mathbb{C}\backslash[0,1)$ for l$q$J and l$q$L cases).
The $n$ zeros of $P_{\cD,n}(\eta;\bm{\lambda})$ in the physical
region are simple and we write them as $\eta^{(n)}_j$ ($j=1,2,\ldots,n$,
$\eta^{(n)}_1<\eta^{(n)}_2<\cdots<\eta^{(n)}_n$) and set $x^{(n)}_j$ as
$\eta^{(n)}_j=\eta(x^{(n)}_j;\bm{\lambda}+M\tilde{\bm{\delta}})$
($\Rightarrow$ $x^{(n)}_1<x^{(n)}_2<\cdots<x^{(n)}_ n$).
Let us define $\bar{x}^{(n)}_j\in\cX$ as $\bar{x}^{(n)}_j=[x^{(n)}_j]$,
where $[a]$ denotes the greatest integer not exceeding $a$.
Then we have $x^{(n)}_j+1<x^{(n)}_{j+1}$
($\Rightarrow$ $\bar{x}^{(n)}_j+1\leq\bar{x}^{(n)}_{j+1}$
$\Rightarrow$ $\check{P}_{\cD,n}(x;\bm{\lambda})$ changes its sign $n$
times in $\cX$)
and the interlacing property
$\bar{x}^{(n+1)}_j\leq\bar{x}^{(n)}_j\leq\bar{x}^{(n+1)}_{j+1}$.
We remark that the property $x^{(n+1)}_j<x^{(n)}_j<x^{(n+1)}_{j+1}$
may not hold.
These properties can be verified by numerical calculation.
The eigenvectors and eigenvalues of $\cH_{\cD}$ are given by
$\phi_{\cD\,n}(x)$ \eqref{phiDndef}, 
\begin{equation}
  \sum_{y\in\cX}\cH_{\cD}(\bm{\lambda})_{x,y}\phi_{\cD\,n}(y;\bm{\lambda})
  =\cE_n(\bm{\lambda})\phi_{\cD\,n}(x;\bm{\lambda})\ \ (n,x\in\cX),
  \label{HDphiDn=}
\end{equation}
where the energy eigenvalues $\cE_n$ satisfy
\begin{equation}
  0=\cE_0(\bm{\lambda})<\cE_1(\bm{\lambda})<\cE_2(\bm{\lambda})<\cdots.
  \label{E0<E1<}
\end{equation}
Since the Hamiltonian $\cH_{\cD}$ is a real symmetric matrix,
its eigenvectors $\phi_{\cD\,n}(x)$ are orthogonal, which gives the
orthogonality relations for $\check{P}_{\cD,n}(x)$ \eqref{PDndef}:
\begin{equation}
  \sum_{x\in\cX}\frac{\phi_0(x;\bm{\lambda}+M\tilde{\bm{\delta}})^2}
  {\check{\Xi}_{\cD}(x;\bm{\lambda})\check{\Xi}_{\cD}(x+1;\bm{\lambda})}
  \check{P}_{\cD,n}(x;\bm{\lambda})\check{P}_{\cD,m}(x;\bm{\lambda})
  =\frac{\delta_{n,m}}
  {d_n(\bm{\lambda})^2\tilde{d}_{\cD,n}(\bm{\lambda})^2}\ \ (n,m\in\cX).
  \label{orthoPDn}
\end{equation}
The inner product $(f,g)$ of two vectors $f=(f(x))_{x\in\cX}$ and
$g=(g(x))_{x\in\cX}$ is defined by
\begin{equation}
  (f,g)\eqdef\sum_{x\in\cX}f(x)g(x).
\end{equation}
The normalized eigenvector $\hat{\phi}_{\cD\,n}(x)$,
$(\hat{\phi}_{\cD\,n},\hat{\phi}_{\cD\,m})=\delta_{n,m}$,
is given by
\begin{equation}
  \hat{\phi}_{\cD\,n}(x;\bm{\lambda})\eqdef
  \frac{d_n(\bm{\lambda})\tilde{d}_{\cD,n}(\bm{\lambda})}
  {\sqrt{\check{\Xi}_{\cD}(1;\bm{\lambda})}}
  \phi_{\cD\,n}(x;\bm{\lambda}).
\end{equation}
Since $\hat{\phi}_{\cD\,n}(x;\bm{\lambda})$'s are orthonormal and
complete, we have the following relations:
\begin{align}
  \sum_{x\in\cX}\hat{\phi}_{\cD\,n}(x;\bm{\lambda})
  \hat{\phi}_{\cD\,m}(x;\bm{\lambda})
  &=\delta_{n,m}\ \ (n,m\in\cX),
  \label{orthonor}\\
  \sum_{n\in\cX}\hat{\phi}_{\cD\,n}(x;\bm{\lambda})
  \hat{\phi}_{\cD\,n}(y;\bm{\lambda})
  &=\delta_{x,y}\ \ (x,y\in\cX).
  \label{comp}
\end{align}
We remark that \eqref{comp} does not hold for $q$M case, see \cite{os34}.
The spectral representation of $\cH_{\cD}$ is given by
\begin{equation}
  \cH_{\cD}(\bm{\lambda})_{x,y}
  =\sum_{n\in\cX}\cE_n(\bm{\lambda})
  \hat{\phi}_{\cD\,n}(x;\bm{\lambda})\hat{\phi}_{\cD\,n}(y;\bm{\lambda}).
\end{equation}

The similarity transformed Hamiltonian $\wtcH_{\cD}$ is defined by a similarity
transformation in terms of a diagonal matrix
$\text{diag}(\psi_{\cD}(0),\psi_{\cD}(1),\psi_{\cD}(2),\ldots)$,
\begin{align}
  &\wtcH_{\cD}(\bm{\lambda})
  =\bigl(\wtcH_{\cD}(\bm{\lambda})_{x,y}\bigr)_{x,y\in\cX}\,,\\
  &\wtcH_{\cD}(\bm{\lambda})_{x,y}
  \eqdef\psi_{\cD}(x;\bm{\lambda})^{-1}
  \cH_{\cD}(\bm{\lambda})_{x,y}\,\psi_{\cD}(y;\bm{\lambda}),\n
  &\phantom{\wtcH_{\cD}(\bm{\lambda})_{x,y}}\,
  =\bigl(B_{\cD}(x;\bm{\lambda})+D_{\cD}(x;\bm{\lambda})\bigr)\delta_{x,y}\\
  &\phantom{\wtcH_{\cD}(\bm{\lambda})_{x,y}=}
  -B(x;\bm{\lambda}+M\tilde{\bm{\delta}})
  \frac{\check{\Xi}_{\cD}(x;\bm{\lambda})}
  {\check{\Xi}_{\cD}(x+1;\bm{\lambda})}\delta_{x+1,y}
  -D(x;\bm{\lambda}+M\tilde{\bm{\delta}})
  \frac{\check{\Xi}_{\cD}(x+1;\bm{\lambda})}
  {\check{\Xi}_{\cD}(x;\bm{\lambda})}\delta_{x-1,y}\,,
  \nonumber
\end{align}
where $\sqrt{B(x;\bm{\lambda})}\,\phi_0(x;\bm{\lambda})
=\sqrt{D(x+1;\bm{\lambda})}\,\phi_0(x+1;\bm{\lambda})$ is used.
The eigenvectors of $\wtcH_{\cD}$ are given by the multi-indexed polynomials
$\check{P}_{\cD,n}(x)$,
\begin{equation}
  \sum_{y\in\cX}\wtcH_{\cD}
  (\bm{\lambda})_{x,y}\check{P}_{\cD,n}(y;\bm{\lambda})
  =\cE_n(\bm{\lambda})\check{P}_{\cD,n}(x;\bm{\lambda})
  \ \ (n,x\in\cX).
  \label{tHDPDn=}
\end{equation}
As mentioned in \S\,\ref{sec:intro}, $\sum_{x\in\cX}\wtcH_{\cD\,x,y}$ does not
vanish (is not a constant) in general.

Let us consider other similarity transformed Hamiltonians. By similarity
transforming $\wtcH_{\cD}(\bm{\lambda})$ in terms of a diagonal matrix
$\text{diag}(\check{\Xi}_{\cD}(0;\bm{\lambda}+\bm{\delta}),
\check{\Xi}_{\cD}(1;\bm{\lambda}+\bm{\delta}),\ldots)$,
we define a matrix $\wtcH'_{\cD}$ as follows:
\begin{align}
  &\wtcH'_{\cD}(\bm{\lambda})
  =\bigl(\wtcH'_{\cD}(\bm{\lambda})_{x,y}\bigr)_{x,y\in\cX}\,,\\
  &\wtcH'_{\cD}(\bm{\lambda})_{x,y}\eqdef
  \check{\Xi}_{\cD}(x;\bm{\lambda}+\bm{\delta})^{-1}
  \wtcH_{\cD}(\bm{\lambda})_{x,y}\,
  \check{\Xi}_{\cD}(y;\bm{\lambda}+\bm{\delta})\n
  &\phantom{\wtcH'_{\cD}(\bm{\lambda})_{x,y}}\,
  =\bigl(B_{\cD}(x;\bm{\lambda})+D_{\cD}(x;\bm{\lambda})\bigr)\delta_{x,y}
  -B_{\cD}(x;\bm{\lambda})\delta_{x+1,y}
  -D_{\cD}(x;\bm{\lambda})\delta_{x-1,y}\,.
\end{align}
The eigenvectors of $\wtcH'_{\cD}$
are given by
\begin{align}
  &\sum_{y\in\cX}\wtcH'_{\cD}(\bm{\lambda})_{x,y}
  \check{R}_{\cD,n}(y;\bm{\lambda})
  =\cE_n(\bm{\lambda})\check{R}_{\cD,n}(x;\bm{\lambda})\ \ (n,x\in\cX),
  \label{tHD'RDn=}\\
  &\check{R}_{\cD,n}(x;\bm{\lambda})\eqdef
  \frac{\check{P}_{\cD,n}(x;\bm{\lambda})}{\check{P}_{\cD,0}(x;\bm{\lambda})},
\end{align}
where the property \eqref{PD0=XiD} is used.
We remark that \eqref{tHD'RDn=} with $n=0$ gives
\begin{equation}
  \sum_{y\in\cX}\wtcH'_{\cD}(\bm{\lambda})_{x,y}=0\ \ (x\in\cX),
  \label{sumytHD'xy=0}
\end{equation}
by $\check{R}_{\cD,0}(x)=1$ and $\cE_0=0$.
The orthogonality relations for $\check{R}_{\cD,n}(x)$ are
\begin{equation}
  \sum_{x\in\cX}\hat{\phi}_{\cD\,0}(x;\bm{\lambda})^2
  \check{R}_{\cD,n}(x;\bm{\lambda})\check{R}_{\cD,m}(x;\bm{\lambda})
  =\frac{d_0(\bm{\lambda})^2\tilde{d}_{\cD,0}(\bm{\lambda})^2}
  {d_n(\bm{\lambda})^2\tilde{d}_{\cD,n}(\bm{\lambda})^2}\delta_{n,m}
  \ \ (n,m\in\cX).
  \label{orthoRDn}
\end{equation}

Next, by similarity transforming $\wtcH'_{\cD}(\bm{\lambda})$ in terms of a
diagonal matrix
$\text{diag}(\phi_{\cD\,0}(0;\bm{\lambda})^{-2},$
$\phi_{\cD\,0}(1;\bm{\lambda})^{-2},\ldots)$,
we define a matrix $\mathcal{G}_{\cD}$ as follows:
\begin{align}
  &\mathcal{G}_{\cD}(\bm{\lambda})
  =\bigl(\mathcal{G}_{\cD}(\bm{\lambda})_{x,y}\bigr)_{x,y\in\cX}\,,\\
  &\mathcal{G}_{\cD}(\bm{\lambda})_{x,y}\eqdef
  \phi_{\cD\,0}(x;\bm{\lambda})^2
  \,\wtcH'_{\cD}(\bm{\lambda})_{x,y}\,
  \phi_{\cD\,0}(y;\bm{\lambda})^{-2}\n
  &\phantom{\mathcal{G}_{\cD}(\bm{\lambda})_{x,y}}\,
  =\bigl(B_{\cD}(x;\bm{\lambda})+D_{\cD}(x;\bm{\lambda})\bigr)\delta_{x,y}
  -D_{\cD}(x+1;\bm{\lambda})\delta_{x+1,y}
  -B_{\cD}(x-1;\bm{\lambda})\delta_{x-1,y}\n
  &\phantom{\mathcal{G}_{\cD}(\bm{\lambda})_{x,y}}\,
  =\wtcH'_{\cD}(\bm{\lambda})_{y,x}\,,
\end{align}
namely $\mathcal{G}_{\cD}(\bm{\lambda})={}^t\wtcH'_{\cD}(\bm{\lambda})$.
The eigenvectors of $\mathcal{G}_{\cD}={}^t\wtcH'_{\cD}$ are given by
\begin{align}
  &\sum_{y\in\cX}\wtcH'_{\cD}(\bm{\lambda})_{y,x}\,
  \Phi_{\cD,n}(y;\bm{\lambda})
  =\cE_n(\bm{\lambda})\Phi_{\cD,n}(x;\bm{\lambda})\ \ (n,x\in\cX),
  \label{ttHD'PhiDn=}\\
  &\Phi_{\cD,n}(x;\bm{\lambda})
  \eqdef\phi_{\cD\,0}(x;\bm{\lambda})^2\check{R}_{\cD,n}(x;\bm{\lambda})
  =\phi_{\cD\,0}(x;\bm{\lambda})\phi_{\cD\,n}(x;\bm{\lambda}).
  \label{PhiDndef}
\end{align}

For $M=0$ case ($\cD=\emptyset$, $\check{\Xi}_{\cD}(x)=1$),
the deformed system reduces to the original system,
$\cH_{\cD}=\cH$,
$\wtcH_{\cD}=\wtcH$,
$\wtcH'_{\cD}=\wtcH'$,
$\phi_{\cD\,n}(x)=\phi_n(x)$,
$\check{P}_{\cD,n}(x)=\check{P}_n(x)$,
$\sum_{y\in\cX}\cH_{x,y}\phi_n(y)=\cE_n\phi_n(x)$,
$\sum_{y\in\cX}\wtcH_{x,y}\check{P}_n(y)=\cE_n\check{P}_n(x)$.
We remark that $\wtcH'$ and $\wtcH$ are the same, $\wtcH'=\wtcH$.

The deformed rdQM systems ($\cH_{\cD}$) have shape invariance inherited from
the original systems ($\cH$), and we obtain the forward and backward shift
relations for the case-(1) multi-indexed polynomials $\check{P}_{\cD,n}(x)$.
Let us define the shift operators $\cF_{\cD}$ and $\cB_{\cD}$ as follows:
\begin{align}
  \cF_{\cD}(\bm{\lambda})
  &\eqdef\frac{B(0;\bm{\lambda}+M\tilde{\bm{\delta}})}
  {\varphi(x;\bm{\lambda}+M\tilde{\bm{\delta}})
  \check{\Xi}_{\cD}(x+1;\bm{\lambda})}
  \bigl(\check{\Xi}_{\cD}(x+1;\bm{\lambda}+\bm{\delta})
  -\check{\Xi}_{\cD}(x;\bm{\lambda}+\bm{\delta})e^{\frac{d}{dx}}\bigr),
  \label{cFDdef}\\
  \cB_{\cD}(\bm{\lambda})
  &\eqdef\frac{1}{B(0;\bm{\lambda}+M\tilde{\bm{\delta}})
  \check{\Xi}_{\cD}(x;\bm{\lambda}+\bm{\delta})}
  \label{cBDdef}\\
  &\quad\times\Bigl(B(x;\bm{\lambda}+M\tilde{\bm{\delta}})
  \check{\Xi}_{\cD}(x;\bm{\lambda})
  -D(x;\bm{\lambda}+M\tilde{\bm{\delta}})
  \check{\Xi}_{\cD}(x+1;\bm{\lambda})e^{-\frac{d}{dx}}\Bigr)
  \varphi(x;\bm{\lambda}+M\tilde{\bm{\delta}}).
  \nonumber
\end{align}
Then, the forward and backward shift relations are given by
\begin{align}
  &\cF_{\cD}(\bm{\lambda})\check{P}_{\cD,n}(x;\bm{\lambda})
  =\cE_n(\bm{\lambda})\check{P}_{\cD,n-1}(x;\bm{\lambda}+\bm{\delta})
  \ \ (n\in\cX),\\
  &\cB_{\cD}(\bm{\lambda})\check{P}_{\cD,n-1}(x;\bm{\lambda}+\bm{\delta})
  =\check{P}_{\cD,n}(x;\bm{\lambda})
  \ \ (n\in\cX\backslash\{0\}),
\end{align}
for $x\in\mathbb{R}$.

\section{Birth and Death Processes}
\label{sec:BD}

In this section, based on the multi-indexed orthogonal polynomials in
\S\,\ref{sec:miop}, we present exactly solvable BD processes with continuous
and discrete times.
The choices of matrices $L^{\text{BD}}_{\cD}$ and $L^{\text{dBD}}_{\cD}$ are
new results, and other calculations are the same as \cite{s09,s21}.

\subsection{Continuous time BD processes}
\label{sec:cBD}

We treat the multi-indexed orthogonal polynomials considered in
\S\,\ref{sec:miop} except for $q$M type.
Let us consider the following continuous time BD process \cite{s09}:
\begin{equation}
  \frac{\partial}{\partial t}\cP(x;t)
  =\sum_{y\in\cX}L^{\text{BD}}_{\cD\ x,y}\cP(y;t)\ \ (x\in\cX).
  \label{BDeq}
\end{equation}
Here $\cP(x;t)$ ($x\in\cX$, $t\in\mathbb{R}$) is
the probability distribution at the continuous time $t$ ($x$: population)
satisfying
\begin{equation}
  \cP(x;t)\ge0,\quad\sum_{x\in\cX}\cP(x;t)=1,
\end{equation}
and the matrix $L^{\text{BD}}_{\cD}$ is given by
\begin{align}
  &L^{\text{BD}}_{\cD}(\bm{\lambda})
  =\bigl(L^{\text{BD}}_{\cD}(\bm{\lambda})_{x,y}\bigr)_{x,y\in\cX}\,,\quad
  L^{\text{BD}}_{\cD}(\bm{\lambda})\eqdef-{}^t\wtcH'_{\cD}(\bm{\lambda}),
  \label{LBDDdef}\\
  &L^{\text{BD}}_{\cD}(\bm{\lambda})_{x,y}
  =-\bigl(B_{\cD}(x;\bm{\lambda})+D_{\cD}(x;\bm{\lambda})\bigr)\delta_{x,y}
  +D_{\cD}(x+1;\bm{\lambda})\delta_{x+1,y}
  +B_{\cD}(x-1;\bm{\lambda})\delta_{x-1,y}\,.
  \label{LBDDxy}
\end{align}
{}From \eqref{BD,DD>0}--\eqref{BD,DD>0inf}, we have
$L^{\text{BD}}_{\cD\ x,x-1}>0$, $L^{\text{BD}}_{\cD\ x,x+1}>0$ and
$L^{\text{BD}}_{\cD\ x,x}<0$.
The potential functions $B_{\cD}(x)$ and $D_{\cD}(x)$ are interpreted as the
birth and death rates, respectively.
The property \eqref{sumytHD'xy=0} gives
\begin{equation}
  \sum_{x\in\cX}L^{\text{BD}}_{\cD}(\bm{\lambda})_{x,y}=0\ \ (y\in\cX),
  \label{sumxLBDxy=0}
\end{equation}
and this ensures the conservation of probability:
\begin{equation*}
  \frac{\partial}{\partial t}\sum_{x\in\cX}\cP(x;t)
  =\sum_{x\in\cX}\frac{\partial}{\partial t}\cP(x;t)
  =\sum_{x\in\cX}\sum_{y\in\cX}L^{\text{BD}}_{\cD\ x,y}\cP(y;t)
  =\sum_{y\in\cX}\cP(y;t)\sum_{x\in\cX}L^{\text{BD}}_{\cD\ x,y}=0,
\end{equation*}
which gives $\sum_{x\in\cX}\cP(x;t)=1$ for all time.
{}From \eqref{ttHD'PhiDn=}--\eqref{PhiDndef}, the eigenvectors of
$L^{\text{BD}}_{\cD}$ are given by
\begin{equation}
  \sum_{y\in\cX}L^{\text{BD}}_{\cD}(\bm{\lambda})_{x,y}\,
  \phi_{\cD\,0}(y;\bm{\lambda})\phi_{\cD\,n}(y;\bm{\lambda})
  =-\cE_n(\bm{\lambda})
  \phi_{\cD\,0}(x;\bm{\lambda})\phi_{\cD\,n}(x;\bm{\lambda})\ \ (n,x\in\cX),
  \label{LBDDeq}
\end{equation}
and the spectral representation of $L^{\text{BD}}_{\cD}$ is given by
\begin{equation}
  L^{\text{BD}}_{\cD}(\bm{\lambda})_{x,y}
  =-\hat{\phi}_{\cD\,0}(x;\bm{\lambda})
  \Bigl(\sum_{n\in\cX}\cE_n(\bm{\lambda})
  \hat{\phi}_{\cD\,n}(x;\bm{\lambda})\hat{\phi}_{\cD\,n}(y;\bm{\lambda})\Bigr)
  \hat{\phi}_{\cD\,0}(y;\bm{\lambda})^{-1}.
\end{equation}

Let us consider two topics: (\romannumeral1) initial value problem,
(\romannumeral2) transition probability from $y$ to $x$.\\
{\bf (\romannumeral1) initial value problem} :
Given an arbitrary initial probability distribution $\cP(x;0)$,
find the probability distribution at a later time $t$, $\cP(x;t)$.
Since $\hat{\phi}_{\cD\,n}(x)$'s are orthonormal and complete,
$\cP(x;0)$ can be expanded as
\begin{equation}
  \cP(x;0)=\hat{\phi}_{\cD\,0}(x;\bm{\lambda})
  \sum_{n\in\cX}c_n\hat{\phi}_{\cD\,n}(x;\bm{\lambda}),
  \ \ c_n=\bigl(\hat{\phi}_{\cD\,n}(x;\bm{\lambda}),
  \hat{\phi}_{\cD\,0}(x;\bm{\lambda})^{-1}\cP(x;0)\bigr),
\end{equation}
and we have $c_0=\sum_{x\in\cX}\cP(x;0)=1$.
Then $\cP(x;t)$ is given by
\begin{equation}
  \cP(x;t)=\hat{\phi}_{\cD\,0}(x;\bm{\lambda})
  \sum_{n\in\cX}c_n\,e^{-\cE_n(\bm{\lambda})t}\,
  \hat{\phi}_{\cD\,n}(x;\bm{\lambda})\ \ (t\ge0),
  \label{Pxtsol}
\end{equation}
because the right hand side of \eqref{Pxtsol} satisfies the same differential
equation \eqref{BDeq},
\begin{align*}
  &\quad\frac{\partial}{\partial t}\Bigl(\hat{\phi}_{\cD\,0}(x)
  \sum_{n\in\cX}c_n\,e^{-\cE_nt}\,\hat{\phi}_{\cD\,n}(x)\Bigr)
  =\sum_{n\in\cX}c_n\,e^{-\cE_nt}\bigl(-\cE_n\,
  \hat{\phi}_{\cD\,0}(x)\hat{\phi}_{\cD\,n}(x)\bigr)\\
  &=\sum_{n\in\cX}c_n\,e^{-\cE_nt}
  \sum_{y\in\cX}L^{\text{BD}}_{\cD\ x,y}\,
  \hat{\phi}_{\cD\,0}(y)\hat{\phi}_{\cD\,n}(y)
  =\sum_{y\in\cX}L^{\text{BD}}_{\cD\ x,y}\Bigl(\hat{\phi}_{\cD\,0}(y)
  \sum_{n\in\cX}c_n\,e^{-\cE_nt}\hat{\phi}_{\cD\,n}(y)\Bigr),
\end{align*}
and becomes $\cP(x;0)$ for $t=0$.
We remark that $\hat{\phi}_{\cD\,0}(x)^2$,
\begin{equation}
  \hat{\phi}_{\cD\,0}(x;\bm{\lambda})^2
  =d_n(\bm{\lambda})^2\tilde{d}_{\cD,n}(\bm{\lambda})^2
  \frac{\check{\Xi}_{\cD}(x;\bm{\lambda}+\bm{\delta})^2}
  {\check{\Xi}_{\cD}(x;\bm{\lambda})\check{\Xi}_{\cD}(x+1;\bm{\lambda})}
  \phi_0(x;\bm{\lambda}+M\tilde{\bm{\delta}})^2,
  \label{hphiD0^2}
\end{equation}
is a stationary probability distribution, because the initial condition
$\cP(x;0)=\hat{\phi}_{\cD\,0}(x)^2$ gives $c_n=\delta_{n,0}$.
In the $t\to\infty$ limit of \eqref{Pxtsol}, $\cP(x;t)$ approaches
the stationary probability distribution,
\begin{equation}
  \lim_{t\to\infty}\cP(x;t)=\hat{\phi}_{\cD\,0}(x;\bm{\lambda})^2,
\end{equation}
by \eqref{E0<E1<}.

\noindent
{\bf (\romannumeral2) transition probability from $y$ to $x$} :
For a concentrated initial distribution at $y$,
$\cP(x;0)=\delta_{x,y}$, find the transition probability from $y$ to
$x$ at time $t$, $\cP(x,y;t)$.
This transition probability $\cP(x,y;t)$ is given by
\begin{equation}
  \cP(x,y;t)=\hat{\phi}_{\cD\,0}(x;\bm{\lambda})
  \Bigl(\sum_{n\in\cX}e^{-\cE_n(\bm{\lambda})t}\,
  \hat{\phi}_{\cD\,n}(x;\bm{\lambda})\hat{\phi}_{\cD\,n}(y;\bm{\lambda})\Bigr)
  \hat{\phi}_{\cD\,0}(y;\bm{\lambda})^{-1}\ \ (t\ge0).
  \label{Pxytsol}
\end{equation}
For $t=0$, the right hand side of \eqref{Pxytsol} becomes
\begin{equation*}
  \hat{\phi}_{\cD\,0}(x;\bm{\lambda})
  \Bigl(\sum_{n\in\cX}\hat{\phi}_{\cD\,n}(x;\bm{\lambda})
  \hat{\phi}_{\cD\,n}(y;\bm{\lambda})\Bigr)
  \hat{\phi}_{\cD\,0}(y;\bm{\lambda})^{-1}
  =\hat{\phi}_{\cD\,0}(x;\bm{\lambda})\delta_{x,y}\,
  \hat{\phi}_{\cD\,0}(y;\bm{\lambda})^{-1}
  =\delta_{x,y}.
\end{equation*}
The expression \eqref{Pxytsol} satisfies the Chapman-Kolmogorov equation
\begin{equation}
  \cP(x,y;t)=\sum_{z\in\cX}\cP(x,z;t-t')\cP(z,y;t')\ \ (0\leq t'\leq t),
  \label{CK}
\end{equation}
because the right hand side of \eqref{CK} becomes
\begin{align*}
  &\quad\hat{\phi}_{\cD\,0}(x)\sum_{n\in\cX}\sum_{m\in\cX}
  e^{-\cE_n(t-t')}e^{-\cE_mt'}
  \hat{\phi}_{\cD\,n}(x)\Bigl(\sum_{z\in\cX}
  \hat{\phi}_{\cD\,n}(z)\hat{\phi}_{\cD\,m}(z)\Bigr)
  \hat{\phi}_{\cD\,m}(y)\hat{\phi}_{\cD\,0}(y)^{-1}\\
  &=\hat{\phi}_{\cD\,0}(x)\sum_{n\in\cX}\sum_{m\in\cX}
  e^{-\cE_n(t-t')}e^{-\cE_mt'}\hat{\phi}_{\cD\,n}(x)
  \delta_{n,m}\hat{\phi}_{\cD\,m}(y)\hat{\phi}_{\cD\,0}(y)^{-1}\\
  &=\hat{\phi}_{\cD\,0}(x)\sum_{n\in\cX}e^{-\cE_nt}\,
  \hat{\phi}_{\cD\,n}(x)\hat{\phi}_{\cD\,n}(y)\hat{\phi}_{\cD\,0}(y)^{-1}
  =\cP(x,y;t).
\end{align*}
In the $t\to\infty$ limit of \eqref{Pxytsol}, $\cP(x,y;t)$ approaches
the stationary probability distribution,
\begin{equation}
  \lim_{t\to\infty}\cP(x,y;t)=\hat{\phi}_{\cD\,0}(x;\bm{\lambda})^2,
\end{equation}
by \eqref{E0<E1<}.

The repeated continuous time BD processes can be obtained from the discrete
time versions, see \S\,\ref{sec:dBDrepeat}.

\subsection{Discrete time BD processes}
\label{sec:dBD}

We treat the multi-indexed orthogonal polynomial of finite type
(H, R, dH, dq$q$K, $q$H, q$q$K, a$q$K, $q$R and d$q$H).
Let us consider the following discrete time BD process (Markov chain)
\cite{s21}:
\begin{equation}
  \cP(x;\ell+1)=\sum_{y\in\cX}L^{\text{dBD}}_{\cD\ x,y}\cP(y;\ell)
  \ \ (x\in\cX).
  \label{dBDeq}
\end{equation}
Here $\cP(x;\ell)$ ($x\in\cX$, $\ell\in\mathbb{Z}$) is
the probability distribution at the discrete time $\ell$ ($x$: state)
satisfying
\begin{equation}
  \cP(x;\ell)\ge0,\quad\sum_{x\in\cX}\cP(x;\ell)=1,
\end{equation}
and the matrix $L^{\text{dBD}}_{\cD}$ is given by ($I$: identity matrix)
\begin{align}
  &L^{\text{dBD}}_{\cD}(\bm{\lambda})
  =\bigl(L^{\text{dBD}}_{\cD}(\bm{\lambda})_{x,y}\bigr)_{x,y\in\cX}\,,\quad
  L^{\text{dBD}}_{\cD}(\bm{\lambda})\eqdef
  I+t_{\text{S}}L^{\text{BD}}_{\cD}(\bm{\lambda}),
  \label{LdBDDdef}\\
  &L^{\text{dBD}}_{\cD}(\bm{\lambda})_{x,y}
  =\bigl(1-t_{\text{S}}\bigl(B_{\cD}(x;\bm{\lambda})
  +D_{\cD}(x;\bm{\lambda})\bigr)\bigr)\delta_{x,y}\n
  &\phantom{L^{\text{dBD}}_{\cD}(\bm{\lambda})_{x,y}=}
  +t_{\text{S}}D_{\cD}(x+1;\bm{\lambda})\delta_{x+1,y}
  +t_{\text{S}}B_{\cD}(x-1;\bm{\lambda})\delta_{x-1,y}\,,
  \label{LdBDDxy}
\end{align}
and the time scale parameter $t_{\text{S}}$ is a positive constant satisfying
the following condition:
\begin{equation}
  t_{\text{S}}\cdot\max_{x\in\cX}
  \big(B_{\cD}(x;\bm{\lambda})+D_{\cD}(x;\bm{\lambda})\bigr)<1.
  \label{tScond}
\end{equation}
{}From \eqref{BD,DD>0}--\eqref{BD,DD>0inf} and \eqref{tScond}, we have
$L^{\text{dBD}}_{\cD\ x,x+1}>0$, $L^{\text{dBD}}_{\cD\ x,x-1}>0$
and $L^{\text{dBD}}_{\cD\ x,x}>0$.
Thus $L^{\text{dBD}}_{\cD}$ is a non-negative tri-diagonal matrix.
The property \eqref{sumxLBDxy=0} gives
\begin{equation}
  \sum_{x\in\cX}L^{\text{dBD}}_{\cD\ x,y}=1\ \ (y\in\cX),
  \label{sumxLdBDxy=1}
\end{equation}
and this ensures the conservation of probability:
\begin{equation*}
  \sum_{x\in\cX}\cP(x;\ell+1)
  =\sum_{x\in\cX}\sum_{y\in\cX}L^{\text{dBD}}_{\cD\ x,y}\cP(y;\ell)
  =\sum_{y\in\cX}\cP(y;\ell)\sum_{x\in\cX}L^{\text{dBD}}_{\cD\ x,y}
  =\sum_{y\in\cX}\cP(y;\ell)=1.
\end{equation*}
{}From \eqref{LBDDeq}, the eigenvectors of $L^{\text{dBD}}_{\cD}$ are given by
\begin{align}
  &\sum_{y\in\cX}L^{\text{dBD}}_{\cD}(\bm{\lambda})_{x,y}\,
  \phi_{\cD\,0}(y;\bm{\lambda})\phi_{\cD\,n}(y;\bm{\lambda})
  =\kappa_n(\bm{\lambda})
  \phi_{\cD\,0}(x;\bm{\lambda})\phi_{\cD\,n}(x;\bm{\lambda})\ \ (n,x\in\cX),
  \label{LdBDDeq}\\
  &\kappa_n(\bm{\lambda})\eqdef 1-t_{\text{S}}\,\cE_n(\bm{\lambda}),
\end{align}
and the spectral representation of $L^{\text{dBD}}_{\cD}$ is given by
\begin{equation}
  L^{\text{BD}}_{\cD}(\bm{\lambda})_{x,y}
  =\hat{\phi}_{\cD\,0}(x;\bm{\lambda})\Bigl(\sum_{n\in\cX}\kappa_n(\bm{\lambda})
  \hat{\phi}_{\cD\,n}(x;\bm{\lambda})\hat{\phi}_{\cD\,n}(y;\bm{\lambda})\Bigr)
  \hat{\phi}_{\cD\,0}(y;\bm{\lambda})^{-1}.
\end{equation}
{}From \eqref{E0<E1<}, eigenvalues $\kappa_n$ satisfy
\begin{equation}
  1=\kappa_0(\bm{\lambda})>\kappa_1(\bm{\lambda})>\kappa_2(\bm{\lambda})
  >\cdots>-1,
  \label{kappa0>kappa1>}
\end{equation}
because the Perron-Frobenius theorem implies $-1\leq\kappa_n\leq 1$ and
$\kappa_n=-1$ is excluded by $\{\kappa_n|n\in\cX\}\neq\{-\kappa_n|n\in\cX\}$
(or by tuning (decreasing) $t_{\text{S}}$, if necessary.)

Let us consider two topics: (\romannumeral1) initial value problem,
(\romannumeral2) transition probability from $y$ to $x$.\\
{\bf (\romannumeral1) initial value problem} :
Given an arbitrary initial probability distribution $\cP(x;0)$,
find the probability distribution at a later time $\ell$, $\cP(x;\ell)$.
Since $\hat{\phi}_{\cD\,n}(x)$'s are orthonormal and complete,
$\cP(x;0)$ can be expanded as
\begin{equation}
  \cP(x;0)=\hat{\phi}_{\cD\,0}(x;\bm{\lambda})
  \sum_{n\in\cX}c_n\hat{\phi}_{\cD\,n}(x;\bm{\lambda}),
  \ \ c_n=\bigl(\hat{\phi}_{\cD\,n}(x;\bm{\lambda}),
  \hat{\phi}_{\cD\,0}(x;\bm{\lambda})^{-1}\cP(x;0)\bigr),
\end{equation}
and we have $c_0=\sum_{x\in\cX}\cP(x;0)=1$.
Then $\cP(x;\ell)$ is given by
\begin{equation}
  \cP(x;\ell)=\hat{\phi}_{\cD\,0}(x;\bm{\lambda})
  \sum_{n\in\cX}c_n\kappa_n(\bm{\lambda})^{\ell}\,
  \hat{\phi}_{\cD\,n}(x;\bm{\lambda})\ \ (\ell\in\mathbb{Z}_{\ge0}),
  \label{dPxtsol}
\end{equation}
because the right hand side of \eqref{dPxtsol} satisfies the same difference
equation \eqref{dBDeq},
\begin{align*}
  &\quad\sum_{y\in\cX}L^{\text{dBD}}_{\cD\ x,y}\Bigl(\hat{\phi}_{\cD\,0}(y)
  \sum_{n\in\cX}c_n\kappa_n^{\ell}\,\hat{\phi}_{\cD\,n}(y)\Bigr)
  =\sum_{n\in\cX}c_n\kappa_n^{\ell}\sum_{y\in\cX}L^{\text{dBD}}_{\cD\ x,y}\,
  \hat{\phi}_{\cD\,0}(y)\hat{\phi}_{\cD\,n}(y)\\
  &=\sum_{n\in\cX}c_n\kappa_n^{\ell}\kappa_n\,
  \hat{\phi}_{\cD\,0}(x)\hat{\phi}_{\cD\,n}(x)
  =\hat{\phi}_{\cD\,0}(x)
  \sum_{n\in\cX}c_n\kappa_n^{\ell+1}\,\hat{\phi}_{\cD\,n}(x),
\end{align*}
and becomes $\cP(x;0)$ for $\ell=0$.
We remark that $\hat{\phi}_{\cD\,0}(x)^2$,
\begin{equation}
  \hat{\phi}_{\cD\,0}(x;\bm{\lambda})^2
  =d_n(\bm{\lambda})^2\tilde{d}_{\cD,n}(\bm{\lambda})^2
  \frac{\check{\Xi}_{\cD}(x;\bm{\lambda}+\bm{\delta})^2}
  {\check{\Xi}_{\cD}(x;\bm{\lambda})\check{\Xi}_{\cD}(x+1;\bm{\lambda})}
  \phi_0(x;\bm{\lambda}+M\tilde{\bm{\delta}})^2,
\end{equation}
is a stationary probability distribution, because the initial condition
$\cP(x;0)=\hat{\phi}_{\cD\,0}(x)^2$ gives $c_n=\delta_{n,0}$.
In the $\ell\to\infty$ limit of \eqref{dPxtsol}, $\cP(x;\ell)$
approaches the stationary probability distribution,
\begin{equation}
  \lim_{\ell\to\infty}\cP(x;\ell)=\hat{\phi}_{\cD\,0}(x;\bm{\lambda})^2,
\end{equation}
by \eqref{kappa0>kappa1>}.

\noindent
{\bf (\romannumeral2) transition probability from $y$ to $x$} :
For a concentrated initial distribution at $y$,
$\cP(x;0)=\delta_{x,y}$, find the transition probability from $y$ to
$x$ at time $\ell$, $\cP(x,y;\ell)$.
This transition probability $\cP(x,y;\ell)$ is given by
\begin{equation}
  \cP(x,y;\ell)=\hat{\phi}_{\cD\,0}(x;\bm{\lambda})
  \Bigl(\sum_{n\in\cX}\kappa_n(\bm{\lambda})^{\ell}\,
  \hat{\phi}_{\cD\,n}(x;\bm{\lambda})\hat{\phi}_{\cD\,n}(y;\bm{\lambda})\Bigr)
  \hat{\phi}_{\cD\,0}(y;\bm{\lambda})^{-1}\ \ (\ell\in\mathbb{Z}_{\ge0}).
  \label{dPxytsol}
\end{equation}
For $\ell=0$, the right hand side of \eqref{dPxytsol} becomes
\begin{equation*}
  \hat{\phi}_{\cD\,0}(x;\bm{\lambda})\Bigl(\sum_{n\in\cX}
  \hat{\phi}_{\cD\,n}(x;\bm{\lambda})\hat{\phi}_{\cD\,n}(y;\bm{\lambda})\Bigr)
  \hat{\phi}_{\cD\,0}(y;\bm{\lambda})^{-1}
  =\hat{\phi}_{\cD\,0}(x;\bm{\lambda})\delta_{x,y}\,
  \hat{\phi}_{\cD\,0}(y;\bm{\lambda})^{-1}
  =\delta_{x,y}.
\end{equation*}
The expression \eqref{dPxytsol} satisfies the Chapman-Kolmogorov equation
\begin{equation}
  \cP(x,y;\ell)=\sum_{z\in\cX}\cP(x,z;\ell-\ell')
  \cP(z,y;\ell')\ \ (0\leq\ell'\leq\ell),
  \label{dCK}
\end{equation}
because the right hand side of \eqref{dCK} becomes
\begin{align*}
  &\quad\hat{\phi}_{\cD\,0}(x)
  \sum_{n\in\cX}\sum_{m\in\cX}\kappa_n^{\ell-\ell'}\kappa_m^{\ell'}\,
  \hat{\phi}_{\cD\,n}(x)\Bigl(\sum_{z\in\cX}
  \hat{\phi}_{\cD\,n}(z)\hat{\phi}_{\cD\,m}(z)\Bigr)
  \hat{\phi}_{\cD\,m}(y)\hat{\phi}_{\cD\,0}(y)^{-1}\\
  &=\hat{\phi}_{\cD\,0}(x)
  \sum_{n\in\cX}\sum_{m\in\cX}\kappa_n^{\ell-\ell'}\kappa_m^{\ell'}\,
  \hat{\phi}_{\cD\,n}(x)\delta_{n,m}
  \hat{\phi}_{\cD\,m}(y)\hat{\phi}_{\cD\,0}(y)^{-1}\\
  &=\hat{\phi}_{\cD\,0}(x)
  \sum_{n\in\cX}\kappa_n^{\ell}\,\hat{\phi}_{\cD\,n}(x)
  \hat{\phi}_{\cD\,n}(y)\hat{\phi}_{\cD\,0}(y)^{-1}
  =\cP(x,y;\ell).
\end{align*}
In the $\ell\to\infty$ limit of \eqref{dPxytsol}, $\cP(x,y;\ell)$
approaches the stationary probability distribution,
\begin{equation}
  \lim_{\ell\to\infty}\cP(x,y;\ell)=\hat{\phi}_{\cD\,0}(x;\bm{\lambda})^2,
\end{equation}
by \eqref{kappa0>kappa1>}.

The continuous time BD process can be recovered from the discrete time BD
process by taking $t_{\text{S}}\to 0$ limit.
By setting $\ell\,t_{\text{S}}=t$ and $\cP(x;\ell)=\cP'(x;t)$,
\eqref{dBDeq} is rewritten as
\begin{equation*}
  \frac{\cP'(x;t+t_{\text{S}})-\cP'(x;t)}{t_{\text{S}}}
  =\sum_{y\in\cX}L^{\text{BD}}_{\cD\ x,y}\cP'(y;t).
\end{equation*}
By taking $t_{\text{S}}\to 0$ limit, this equation gives
$\frac{\partial}{\partial t}\cP'(x;t)
=\sum_{y\in\cX}L^{\text{BD}}_{\cD\ x,y}\cP'(y;t)$,
\eqref{BDeq}.

\subsubsection{repeated discrete time BD processes}
\label{sec:dBDrepeat}

Repeated discrete time BD processes (Markov chain) are studied for the
orthogonal polynomials of a discrete variable in the Askey scheme \cite{os39}.
This method can be applied to the multi-indexed orthogonal polynomial cases.

We can show that the $m$-th power of $L^{\text{BD}}_{\cD}$ \eqref{LBDDdef},
$L^{\text{BD}\,m}_{\cD}$ ($m\in\mathbb{Z}_{\ge1}$), has the following form of
the matrix elements,
\begin{equation*}
  (L^{\text{BD}\,m}_{\cD})_{x+k,x}=(-1)^{m-k}a^{(m)}_k(x)\ \ (-m\le k\le m),
  \quad(L^{\text{BD}\,m}_{\cD})_{x,y}=0\ \ (|x-y|>m),
\end{equation*}
where $a^{(m)}_k(x)>0$.
Let us consider the following matrix $X_{\cD}$,
\begin{equation}
  X_{\cD}\eqdef\sum_{j=0}^{m-1}c_jL^{\text{BD}\,m-j}_{\cD},\ \ c_0=1\quad
  \Bigl(\Rightarrow\sum_{x\in\cX}X_{\cD\,x,y}=0,\ X_{\cD\,x,y}=0
  \ (|x-y|>m)\Bigr),
  \label{XDdef}
\end{equation}
where $c_j$ are constants. Its non zero matrix elements are
\begin{align*}
  X_{\cD\,x\pm(m-k),x}&=\sum_{j=0}^kc_j(-1)^{k-j}a^{(m-j)}_{\pm(m-k)}(x)
  \ \ (0\le k\le m-1),\\
  X_{\cD\,x,x}&=\sum_{j=0}^{m-1}c_j(-1)^{m-j}a^{(m-j)}_0(x).
\end{align*}
Starting from $X_{\cD\,x\pm m,x}=a^{(m)}_{\pm m}(x)>0$, we can tune $c_k$
($k=1,\ldots,m-2$ in turn) such that $X_{\cD\,x\pm(m-k),x}>0$, and tune
$c_{m-1}$ such that $X_{\cD\,x\pm 1,x}>0$ and $X_{\cD\,x,x}<0$.
For such chosen weights $\{c_j\}$ and a positive constant $t_{\text{S}}$,
we define a matrix $L^{\text{dBD}(m)}_{\cD}$,
\begin{equation}
  L^{\text{dBD}(m)}_{\cD}\eqdef I+t_{\text{S}}\,X_{\cD},\quad
  t_{\text{S}}\cdot\max\bigl(-X_{\cD\,x,x}\bigr)<1,
\end{equation}
which satisfies
\begin{align}
  &L^{\text{dBD}(m)}_{\cD\ x,y}\ge0\ \ (x,y\in\cX),\quad
  L^{\text{dBD}(m)}_{\cD\ x,y}=0\ \ (|x-y|>m),\n
  &\sum_{x\in\cX}L^{\text{dBD}(m)}_{\cD\ x,y}=1\ \ (y\in\cX).
\end{align}
This gives an exactly solvable Markov chain
\begin{equation}
  \cP(x;\ell+1)=\sum_{y\in\cX}L^{\text{dBD}(m)}_{\cD\ x,y}\cP(y;\ell)
  \ \ (x\in\cX),
  \label{dBD(m)eq}
\end{equation}
and the matrices $L^{\text{dBD}(m)}_{\cD}$'s have common eigenvectors
\begin{align}
  &\sum_{y\in\cX}L^{\text{dBD}(m)}_{\cD}(\bm{\lambda})_{x,y}
  \phi_{\cD\,0}(y;\bm{\lambda})\phi_{\cD\,n}(y;\bm{\lambda})
  =\kappa^{(m)}_n(\bm{\lambda})
  \phi_{\cD\,0}(x;\bm{\lambda})\phi_{\cD\,n}(x;\bm{\lambda})
  \ \ (n,x\in\cX),\\
  &\kappa^{(m)}_n(\bm{\lambda})\eqdef 1+t_{\text{S}}
  \sum_{j=0}^{m-1}(-1)^{m-j}c_j\cE_n(\bm{\lambda})^{m-j}.
\end{align}
The initial value problem and the transition probability from $y$ to $x$ are
solved by the same formulas \eqref{dPxtsol} and \eqref{dPxytsol} with
$\kappa_n$ replaced by $\kappa^{(m)}_n$, respectively.

By taking the $t_{\text{S}}\to 0$ limit (see the paragraph immediately
preceding \S\,\ref{sec:dBDrepeat}), the repeated discrete time BD process
\eqref{dBD(m)eq} gives the repeated continuous time BD process
\begin{equation}
  \frac{\partial}{\partial t}\cP(x;t)
  =\sum_{y\in\cX}L^{\text{BD}(m)}_{\cD\ x,y}\cP(y;t)\ \ (x\in\cX),\quad
  L^{\text{BD}(m)}_{\cD}\eqdef X_{\cD}.
  \label{BD(m)eq}
\end{equation}
The matrix elements $L^{\text{BD}(m)}_{\cD\,x,x\mp k}$ are interpreted as
the birth and death rates for $k$ persons collectively.
The matrices $L^{\text{BD}(m)}_{\cD}$'s have common eigenvectors
\begin{align}
  &\sum_{y\in\cX}L^{\text{BD}(m)}_{\cD}(\bm{\lambda})_{x,y}
  \phi_{\cD\,0}(y;\bm{\lambda})\phi_{\cD\,n}(y;\bm{\lambda})
  =\cE^{(m)}_n(\bm{\lambda})
  \phi_{\cD\,0}(x;\bm{\lambda})\phi_{\cD\,n}(x;\bm{\lambda})
  \ \ (n,x\in\cX),\\
  &\cE^{(m)}_n(\bm{\lambda})\eqdef
  \sum_{j=0}^{m-1}(-1)^{m-j}c_j\cE_n(\bm{\lambda})^{m-j}.
\end{align}
The initial value problem and the transition probability from $y$ to $x$ are
solved by the same formulas \eqref{Pxtsol} and \eqref{Pxytsol} with
$\cE_n$ replaced by $\cE^{(m)}_n$, respectively.
This repeated continuous time BD process is valid for all polynomials in
\S\,\ref{sec:miop} except for $q$M type.

\section{Summary and Comments}
\label{sec:summary}

The case-(1) multi-indexed orthogonal polynomials of a discrete variable
constructed so far are R, $q$R, M, l$q$J and l$q$L types \cite{os26,os35}.
By the same methods in \cite{os26,os35} (with some modification), the case-(1)
multi-indexed orthogonal polynomials of
H, dH, dq$q$K, $q$H, q$q$K, a$q$K, d$q$H and $q$M
types are concretely constructed in \S\,\ref{sec:miop}.

Exactly solvable BD processes are obtained for each orthogonal polynomials of
a discrete variable in the Askey-scheme \cite{s09}, where the matrix
$L^{\text{BD}}$ is given by the similarity transformed Hamiltonian $\wtcH$,
as $L^{\text{BD}}=-{}^t\wtcH$.
For the multi-indexed orthogonal polynomial cases, the choice
$L^{\text{BD}}_{\cD}=-{}^t\wtcH_{\cD}$ does not give the BD processes,
because the conservation of probability is violated.
By considering other similarity transformed Hamiltonian $\wtcH'_{\cD}$, exactly
solvable BD processes are obtained as $L^{\text{BD}}_{\cD}=-{}^t\wtcH'_{\cD}$
in \S\,\ref{sec:BD}.
The discrete time versions and repeated versions are also obtained.
The type $\II$ multi-indexed little $q$-Jacobi and little $q$-Laguerre
polynomials constructed in \cite{lqJII} are the case-(1) polynomials,
but we do not treat them in this paper, because their some expressions are
slightly different from those in \S\,\ref{sec:miop}.
The construction method of the BD process can be applied to them as well.

The construction method of the BD process in \S\,\ref{sec:BD} can also be
applied to the case-(2) polynomials and more general situations.
Let us consider a real symmetric matrix $H$ and assume that its eigenvectors
and eigenvalues are given by
\begin{equation}
  H=(H_{x,y})_{x,y\in\cX},\quad
  \sum_{y\in\cX}H_{x,y}\psi_n(y)=E_n\psi_n(x)\ \ (n,x\in\cX),\quad
  \psi_n(x)\eqdef\psi(x)\check{p}_n(x),
\end{equation}
where $0=E_0<E_1<E_2<\cdots$ (if $E_0\neq0$, consider $H-E_0$).
The vectors $\psi_n(x)$'s are orthogonal and we assume $\check{p}_0(x)\neq0$
($x\in\cX$).
A similarity transformed matrix $\widetilde{H}$ and its eigenvectors are
\begin{align}
  &\widetilde{H}=(\widetilde{H}_{x,y})_{x,y\in\cX},\quad
  \widetilde{H}_{x,y}\eqdef\psi(x)^{-1}H_{x,y}\,\psi(y),\n
  &\sum_{y\in\cX}\widetilde{H}_{x,y}\,\check{p}_n(y)=E_n\,\check{p}_n(x)
  \ \ (n,x\in\cX),
\end{align}
and other similarity transformed matrix $\widetilde{H}'$ and its eigenvectors
are
\begin{align}
  &\widetilde{H}'=(\widetilde{H}'_{x,y})_{x,y\in\cX},\quad
  \widetilde{H}'_{x,y}\eqdef\check{p}_0(x)^{-1}\widetilde{H}_{x,y}
  \,\check{p}_0(y),\n
  &\sum_{y\in\cX}\widetilde{H}'_{x,y}\,\check{r}_n(y)=E_n\check{r}_n(x)
  \ \ (n,x\in\cX),\quad
  \check{r}_n(x)\eqdef\frac{\check{p}_n(x)}{\check{p}_0(x)}.
\end{align}
We remark that this equation with $n=0$ gives
\begin{equation}
  \sum_{y\in\cX}\widetilde{H}'_{x,y}=0\ \ (x\in\cX),
\end{equation}
by $\check{r}_0(x)=1$ and $E_0=0$.
Let us define other similarity transformed matrix $G$,
\begin{equation}
  G=(G_{x,y})_{x,y\in\cX},\quad
  G_{x,y}\eqdef\psi_0(x)^2\widetilde{H}'_{x,y}\,\psi_0(y)^{-2}.
\end{equation}
So we have
\begin{equation}
  \widetilde{H}'_{x,y}=\psi_0(x)^{-1}H_{x,y}\,\psi_0(y),\quad
  G_{x,y}=\psi_0(x)H_{x,y}\,\psi_0(y)^{-1}.
\end{equation}
By $H_{x,y}=H_{y,x}$, we have $G_{x,y}=\widetilde{H}'_{y,x}$,
namely $G={}^t\widetilde{H}'$.
The eigenvectors of $G={}^t\widetilde{H}'$ are given by
\begin{align}
  &\sum_{y\in\cX}\widetilde{H}'_{y,x}\Psi_n(y)=E_n\Psi_n(x)
  \ \ (n,x\in\cX),\\
  &\Psi_n(x)\eqdef\psi_0(x)^2\,\check{r}_n(x)=\psi_0(x)\psi_n(x).
\end{align}
If $\widetilde{H}'_{x,y}\leq0$ ($x\neq y$)
($\Rightarrow$ $\widetilde{H}'_{x,x}>0$), we obtain the BD process with
$L^{\text{BD}}=-{}^t\widetilde{H}'$, whose interaction is not restricted to
the nearest neighbors.
If $\widetilde{H}'_{x,x}$ is bounded, we obtain the discrete time BD process
with $L^{\text{dBD}}=1+t_{\text{S}}(-{}^t\widetilde{H}')$
($t_{\text{S}}\cdot\max_{x\in\cX}\widetilde{H}'_{x,x}<1$).
The Krein-Adler type multi-indexed orthogonal polynomials \cite{os22,addrdQM}
and the multi-indexed orthogonal polynomials obtained by the state adding
Darboux transformations \cite{addrdQM} satisfy the above conditions, and
the exactly solvable BD processes can be obtained.
The multi-indexed orthogonal polynomials studied in \cite{dualmiopqR} are
`ordinary' orthogonal polynomials (namely, satisfy the three term recurrence
relations) and `Krall type' (namely, satisfy $2L$-th order difference equation
($L\geq M+1$)).
If the condition $\widetilde{H}'_{x,y}\leq0$ ($x\neq y$) is checked,
they also give the exactly solvable BD processes, whose interaction range is
$L$ (namely, the state at $x$ interacts with those at $x\pm 1,\ldots,x\pm L$).

Based on the orthogonal polynomials of a discrete variable in the Askey-scheme,
quadratic fermionic oscillator chains are studied, e.g.,
\cite{gvz12,cnv19,bpv24,s24a,s24b}.
Here we comment on their algebraic aspects (not their physical contents).
The construction of exactly solvable quadratic oscillator Hamiltonians is
possible for bosonic oscillators as well as fermionic oscillators.
Let $a_x$ and $a_x^{\dagger}$ ($x\in\cX$) be free oscillators satisfying
\begin{align}
  \text{fermionic}&:\ \{a_x,a_y^{\dagger}\}=\delta_{x,y},
  \ \ \{a_x,a_y\}=\{a_x^{\dagger},a_y^{\dagger}\}=0,\n
  \text{bosonic}&:\ [a_x,a_y^{\dagger}]=\delta_{x,y},
  \ \ [a_x,a_y]=[a_x^{\dagger},a_y^{\dagger}]=0.
  \label{axayd}
\end{align}
For a matrix $A=(A_{x,y})_{x,y}$ ($A_{x,y}\in\mathbb{C}$), let us define an
operator $\hat{O}_A\eqdef\sum_{x,y\in\cX}a_x^{\dagger}A_{x,y}a_y$.
For any two matrices $A$ and $B$, we have
\begin{equation}
  [\hat{O}_A,\hat{O}_B]=\hat{O}_{[A,B]}.
\end{equation}
For a hermitian matrix $A$, let us write $\hat{O}_A$ as $\hat{O}_A=\hat{H}_A$,
which is hermite, $\hat{H}_A^{\dagger}=\hat{H}_A$.
Since any hermitian matrix $A$ is diagonalizable, we have
$U^{\dagger}AU=\text{diag}(\alpha_0,\alpha_1,\alpha_2,\ldots)$
($\alpha_n\in\mathbb{R}$), where $U=(U_{x,n})_{x,n\in\cX}$ is a unitary matrix.
By writing $U_{x,n}=u^{(n)}_x$, we have the following relations:
\begin{equation}
  \sum_{x\in\cX}u^{(n)*}_xu^{(m)}_x=\delta_{n,m}\,,\quad
  \sum_{n\in\cX}u^{(n)}_xu^{(n)*}_y=\delta_{x,y}\,,\quad
  \sum_{n\in\cX}\alpha_nu^{(n)}_xu^{(n)*}_y=A_{x,y}\,.
\end{equation}
Let us define $b_n\eqdef\sum_{x\in\cX}u^{(n)*}_xa_x$ ($n\in\cX$)
($\Rightarrow$ $a_x=\sum_{n\in\cX}u^{(n)}_xb_n$), which are free oscillators
satisfying \eqref{axayd} with the replacement $(a,x,y)\to(b,n,m)$.
Then $\hat{H}_A$ is diagonalized as
$\hat{H}_A=\sum_{n\in\cX}\alpha_nb_n^{\dagger}b_n$, and the partition function
is obtained as
\begin{align}
  \text{fermionic}&:\ \text{Tr}_{\cF}\,e^{-\beta\hat{H}_A}
  =\prod_{n\in\cX}(1+e^{-\beta\alpha_n}),\n
  \text{bosonic}&:\ \text{Tr}_{\cF}\,e^{-\beta\hat{H}_A}
  =\prod_{n\in\cX}(1-e^{-\beta\alpha_n})^{-1},
\end{align}
where $\cF$ is the Fock space.
By choosing $A$ with explicitly known $U$ and $\alpha_n$, we obtain exactly
solvable quadratic oscillator Hamiltonian $\hat{H}_A$.
Exactly solvable quadratic fermionic oscillator Hamiltonians are considered
in \cite{s24a}, based on the 15 orthogonal polynomials of a discrete variable
in the Askey-scheme. This corresponds to $A=\mathcal{H}$, and its multi-indexed
polynomial version $A=\mathcal{H}_{\mathcal{D}}$ is possible.
By using the matrix $K(x,y)$ studied in \cite{os39}, exactly solvable quadratic
fermionic oscillator Hamiltonians are considered in \cite{s24b}.
We think that its multi-indexed polynomial version is difficult.

\section*{Acknowledgements}

I thank the support by Course of Physics, Department of Science.

\bigskip
\appendix
\section{Data for Multi-indexed Orthogonal Polynomials}
\label{app:data}

Here we present data for the case-(1) multi-indexed orthogonal polynomials
of a discrete variable.
After giving our notation and the common quantities in \S\,\ref{sec:common},
we present the basic data for each polynomial in \S\,\ref{sec:eachpoly}.

We have five sinusoidal coordinate $\eta(x)$ in rdQM \cite{os12}:
\begin{align*}
  &\text{finite system}:\left\{
  \begin{array}{rll}
  \text{(\romannumeral1)}:&\eta(x)=x&:\ \text{H,\,K},\\[2pt]
  \text{(\romannumeral2)}:&\eta(x)=x(x+d)&:\ \text{R,\,dH},\\[2pt]
  \text{(\romannumeral3)}:&\eta(x)=1-q^x&:\ \text{dq$q$K},\\[2pt]
  \text{(\romannumeral4)}:&\eta(x)=q^{-x}-1
  &:\ \text{$q$H,\,$q$K,\,q$q$K,\,a$q$K},\\[2pt]
  \text{(\romannumeral5)}:&\eta(x)=(q^{-x}-1)(1-dq^x)
  &:\ \text{$q$R,\,d$q$H,\,d$q$K},
  \end{array}\right.\\
  &\text{semi-infinite system}:\left\{
  \begin{array}{rll}
  \text{(\romannumeral1)}:&\eta(x)=x&:\ \text{M,\,C},\\[2pt]
  \text{(\romannumeral3)}:&\eta(x)=1-q^x&:\ \text{l$q$J,\,l$q$L,\,$q$B},\\[2pt]
  \text{(\romannumeral4)}:&\eta(x)=q^{-x}-1&:\ \text{$q$M,\,ASC\II,\,$q$C},
  \end{array}\right.
\end{align*}
where the abbreviations not mentioned so far are
Krawtchouk (K),
$q$-Krawtchouk ($q$K),
dual $q$-Krawtchouk (d$q$K),
Charlier (C),
$q$-Bessel ($q$B) (= alternative $q$-Charlier),
Al-Salam-Carlitz $\II$ (ASC$\II$),
$q$-Charlier ($q$C).
\ignore{
Here the abbreviations are as follows:
Hahn (H),
Krawtchouk (K),
Racah (R),
dual Hahn (dH),
dual quantum $q$-Krawtchouk (dq$q$K) (which is not treated in \cite{kls}),
$q$-Hahn ($q$H),
$q$-Krawtchouk ($q$K),
quantum $q$-Krawtchouk (q$q$K),
affine $q$-Krawtchouk (a$q$K),
$q$-Racah ($q$R),
dual $q$-Hahn (d$q$H),
dual $q$-Krawtchouk (d$q$K),
Meixner (M),
Charlier (C),
little $q$-Jacobi (l$q$J),
little $q$-Laguerre/Wall (l$q$L),
$q$-Bessel ($q$B) (=alternative $q$-Charlier),
$q$-Meixner ($q$M),
Al-Salam-Carlitz $\II$ (ASC$\II$),
$q$-Charlier ($q$C).
}
The case-(1) multi-indexed orthogonal polynomials were constructed for
R and $q$R \cite{os26}, M, l$q$J and l$q$L \cite{os35}.
The case-(1) type $\II$ multi-indexed l$q$J and l$q$L orthogonal polynomials
were constructed in \cite{lqJII}, but we do not treat them here, because their
some expressions are slightly different.
The case-(1) multi-indexed orthogonal polynomials of
H, dH, dq$q$K, $q$H, q$q$K, a$q$K, d$q$H and $q$M
types are new results.
For other types, we have not found the case-(1) multi-indexed orthogonal
polynomials.

\subsection{Common quantities}
\label{sec:common}

Various quantities depend on a set of parameters
$\bm{\lambda}=(\lambda_1,\lambda_2,\ldots)$ and $q$ ($0<q<1$), and
$q^{\bm{\lambda}}$ stands for
$q^{(\lambda_1,\lambda_2,\ldots)}=(q^{\lambda_1},q^{\lambda_2},\ldots)$.
Their dependence is expressed as $f=f(\bm{\lambda})$ and
$f(x)=f(x;\bm{\lambda})$ if necessary, but $q$-dependence is suppressed.

Definitions of common quantities are as follows \cite{os26,os35}:
\begin{align}
  &\cD\eqdef\{d_1,d_2,\ldots,d_M\}
  \ \ (d_j\in\mathbb{Z}_{\geq1}\text{ : mutually distinct}),\n
  &\qquad\ \text{standard order : }1\leq d_1<d_2<\cdots<d_M,
  \label{stD}\\
  &\ell_{\cD}\eqdef\sum_{j=1}^Md_j-\frac12M(M-1),
  \label{lDdef}\\
  &\check{P}_n(x;\bm{\lambda})\eqdef
  P_n\bigl(\eta(x;\bm{\lambda});\bm{\lambda}\bigr),
  \label{cPndef}\\
  &\check{\xi}_{\text{v}}(x;\bm{\lambda})\eqdef
  \xi_{\text{v}}\bigl(\eta(x;\bm{\lambda});\bm{\lambda}\bigr),\quad
  \check{\xi}_{\text{v}}(x;\bm{\lambda})\eqdef
  \check{P}_{\text{v}}\bigl(x;\mathfrak{t}(\bm{\lambda})\bigr),
  \label{xivdef}\\
  &B'(x;\bm{\lambda})\eqdef B\bigl(x;\mathfrak{t}(\bm{\lambda})\bigr),\quad
  D'(x;\bm{\lambda})\eqdef D\bigl(x;\mathfrak{t}(\bm{\lambda})\bigr),
  \label{B'D'def}\\
  &\cE'_{\text{v}}(\bm{\lambda})\eqdef
  \cE_{\text{v}}\bigl(\mathfrak{t}(\bm{\lambda})\bigr),
  \label{E'vdef}\\
  &\tcE_{\text{v}}(\bm{\lambda})\eqdef
  \alpha(\bm{\lambda})\cE'_{\text{v}}(\bm{\lambda})+\alpha'(\bm{\lambda}),
  \label{Etvdef}\\
  &\phi_0(x;\bm{\lambda})^2\eqdef
  \prod_{y=0}^{x-1}\frac{B(y;\bm{\lambda})}{D(y+1;\bm{\lambda})},
  \ \ \phi_0(x;\bm{\lambda})>0,
  \label{phi0def}\\
  &\tilde{\phi}_0(x;\bm{\lambda})^2\eqdef
  \prod_{y=0}^{x-1}\frac{B'(y;\bm{\lambda})}{D'(y+1;\bm{\lambda})},
  \ \ \tilde{\phi}_0(x;\bm{\lambda})>0
  \ \ \Bigl(\eqref{B'D'def}\Rightarrow\tilde{\phi}_0(x;\bm{\lambda})
  =\phi_0\bigl(x;\mathfrak{t}(\bm{\lambda})\bigr)\Bigr),
  \label{phit0def}\\
  &\nu(x;\bm{\lambda})\eqdef
  \frac{\phi_0(x;\bm{\lambda})}{\tilde{\phi}_0(x;\bm{\lambda})},\quad
  r_j(x_j)=r_j(x_j;\bm{\lambda},M)\eqdef
  \frac{\nu(x_j;\bm{\lambda})}{\nu(x;\bm{\lambda}+M\tilde{\bm{\delta}})},
  \quad x_j\eqdef x+j-1,
  \label{nu,rjdef}\\
  &\varphi(x;\bm{\lambda})\eqdef
  \frac{\eta(x+1;\bm{\lambda})-\eta(x;\bm{\lambda})}{\eta(1;\bm{\lambda})},
  \label{varphidef}\\
  &\varphi_M(x;\bm{\lambda})\eqdef
  \prod_{1\leq j<k\leq M}
  \frac{\eta(x+k-1;\bm{\lambda})-\eta(x+j-1;\bm{\lambda})}
  {\eta(k-j;\bm{\lambda})}
  \quad\bigl(\varphi_0(x)=\varphi_1(x)=1\bigr)\n
  &\phantom{\varphi_M(x;\bm{\lambda})}
  =\prod_{1\leq j<k\leq M}\varphi
  \bigl(x+j-1;\bm{\lambda}+(k-j-1)\bm{\delta}\bigr),
  \label{varphiMdef}\\
  &\mathcal{C}_{\cD}(\bm{\lambda})\eqdef
  \frac{1}{\varphi_M(0;\bm{\lambda})}\prod_{1\leq j<k\leq M}
  \frac{\tcE_{d_j}(\bm{\lambda})-\tcE_{d_k}(\bm{\lambda})}
  {\alpha(\bm{\lambda})B'(j-1;\bm{\lambda})},
  \label{CDdef}\\
  &\mathcal{C}_{\cD,n}(\bm{\lambda})\eqdef
  (-1)^M\mathcal{C}_{\cD}(\bm{\lambda})\tilde{d}_{\cD,n}(\bm{\lambda})^2,
  \label{CDndef}\\
  &\tilde{d}_{\cD,n}(\bm{\lambda})^2
  \eqdef\frac{\varphi_M(0;\bm{\lambda})}{\varphi_{M+1}(0;\bm{\lambda})}
  \prod_{j=1}^M\frac{\cE_n(\bm{\lambda})-\tcE_{d_j}(\bm{\lambda})}
  {\alpha(\bm{\lambda})B'(j-1;\bm{\lambda})},
  \ \ \tilde{d}_{\cD,n}(\bm{\lambda})>0.
  \label{dtDndef}\\
  &\text{W}_{\text{C}}[f_1,f_2,\ldots,f_n](x)
  \eqdef\det\Bigl(f_k(x+j-1)\Bigr)_{1\leq j,k\leq n}
  \ \ \bigl(\text{for }f_i=f_i(x)\bigr),
  \label{WCdef}\\
  &\check{\Xi}_{\cD}(x;\bm{\lambda})\eqdef
  \check{\Xi}_{\cD}\bigl(\eta(x;\bm{\lambda}+(M-1)\tilde{\bm{\delta}});
  \bm{\lambda}\bigr)\eqdef\frac{\text{W}_{\text{C}}
  [\check{\xi}_{d_1},\check{\xi}_{d_2},\ldots,\check{\xi}_{d_M}]
  (x;\bm{\lambda})}
  {\mathcal{C}_{\cD}(\bm{\lambda})\varphi_M(x;\bm{\lambda})},
  \label{XiDdef}\\
  &\check{P}_{\cD,n}(x;\bm{\lambda})\eqdef
  \check{P}_{\cD,n}\bigl(\eta(x;\bm{\lambda}+M\tilde{\bm{\delta}});
  \bm{\lambda}\bigr)
  \eqdef\frac{\text{W}_{\text{C}}
  [\check{\xi}_{d_1},\check{\xi}_{d_2},\ldots,\check{\xi}_{d_M},
  \nu\check{P}_n](x;\bm{\lambda})}
  {\mathcal{C}_{\cD,n}(\bm{\lambda})\varphi_{M+1}(x;\bm{\lambda})
  \nu(x;\bm{\lambda}+M\tilde{\bm{\delta}})}
  \label{PDndef}\\
  &\phantom{\check{P}_{\cD,n}(x;\bm{\lambda})}
  =\mathcal{C}_{\mathcal{D},n}(\bm{\lambda})^{-1}
  \varphi_{M+1}(x;\bm{\lambda})^{-1}\n
  &\phantom{\check{P}_{\cD,n}(x;\bm{\lambda})}
  \quad\times\left|
  \begin{array}{cccc}
  \check{\xi}_{d_1}(x_1)&\cdots&\check{\xi}_{d_M}(x_1)
  &r_1(x_1)\check{P}_n(x_1)\\
  \check{\xi}_{d_1}(x_2)&\cdots&\check{\xi}_{d_M}(x_2)
  &r_2(x_2)\check{P}_n(x_2)\\
  \vdots&\cdots&\vdots&\vdots\\
  \check{\xi}_{d_1}(x_{M+1})&\cdots&\check{\xi}_{d_M}(x_{M+1})
  &r_{M+1}(x_{M+1})\check{P}_n(x_{M+1})\\
  \end{array}\right|,\n
  &B_{\cD}(x;\bm{\lambda})\eqdef
  B(x;\bm{\lambda}+M\tilde{\bm{\delta}})
  \frac{\check{\Xi}_{\cD}(x;\bm{\lambda})}
  {\check{\Xi}_{\cD}(x+1;\bm{\lambda})}
  \frac{\check{\Xi}_{\cD}(x+1;\bm{\lambda}+\bm{\delta})}
  {\check{\Xi}_{\cD}(x;\bm{\lambda}+\bm{\delta})},
  \label{BDdef}\\
  &D_{\cD}(x;\bm{\lambda})\eqdef
  D(x;\bm{\lambda}+M\tilde{\bm{\delta}})
  \frac{\check{\Xi}_{\cD}(x+1;\bm{\lambda})}
  {\check{\Xi}_{\cD}(x;\bm{\lambda})}
  \frac{\check{\Xi}_{\cD}(x-1;\bm{\lambda}+\bm{\delta})}
  {\check{\Xi}_{\cD}(x;\bm{\lambda}+\bm{\delta})},
  \label{DDdef}\\
  &\psi_{\cD}(x;\bm{\lambda})\eqdef
  \sqrt{\check{\Xi}_{\cD}(1;\bm{\lambda})}\,
  \frac{\phi_0(x;\bm{\lambda}+M\tilde{\bm{\delta}})}
  {\sqrt{\check{\Xi}_{\cD}(x;\bm{\lambda})
  \check{\Xi}_{\cD}(x+1;\bm{\lambda})}},
  \label{psiDdef}\\
  &\phi_{\cD\,n}(x;\bm{\lambda})\eqdef
  \psi_{\cD}(x;\bm{\lambda})\check{P}_{\cD,n}(x;\bm{\lambda}),
  \label{phiDndef}\\
  &\check{P}_n(x;\bm{\lambda})=c_n(\bm{\lambda})\eta(x;\bm{\lambda})^n
  +(\text{lower degree terms})
  \ \ (\leftarrow\,\text{def.\ of}\ c_n(\bm{\lambda})),
  \label{cndef}\\
  &\check{\xi}_{\text{v}}(x;\bm{\lambda})
  =\tilde{c}_{\text{v}}(\bm{\lambda})\eta(x;\bm{\lambda})^{\text{v}}
  +(\text{lower degree terms})
  \ \ (\leftarrow\,\text{def.\ of}\ \tilde{c}_{\text{v}}(\bm{\lambda})),
  \label{ctvdef}\\
  &\check{\Xi}_{\cD}(x;\bm{\lambda})=c^{\Xi}_{\cD}(\bm{\lambda})
  \eta\bigl(x;\bm{\lambda}+(M-1)\tilde{\bm{\delta}}\bigr)^{\ell_{\cD}}
  +(\text{lower degree terms})
  \ \ (\leftarrow\,\text{def.\ of}\ c^{\Xi}_{\cD}(\bm{\lambda})),
  \label{cXiDdef}\\
  &\check{P}_{\cD,n}(x;\bm{\lambda})=c^P_{\cD,n}(\bm{\lambda})
  \eta(x;\bm{\lambda}+M\tilde{\bm{\delta}})^{\ell_{\cD}+n}
  +(\text{lower degree terms})
  \ \ (\leftarrow\,\text{def.\ of}\ c^P_{\cD,n}(\bm{\lambda})),\!
  \label{cPDnDdef}\\
  &c_n(\bm{\lambda})=(-1)^n\kappa^{-\binom{n}{2}}
  \prod_{j=1}^n\frac{\cE_n(\bm{\lambda})-\cE_{j-1}(\bm{\lambda})}
  {\eta(j;\bm{\lambda})B(0,\bm{\lambda}+(j-1)\bm{\delta})}
  \ \ \Bigl(\eqref{xivdef}\Rightarrow\tilde{c}_{\text{v}}(\bm{\lambda})
  =c_{\text{v}}\bigl(\mathfrak{t}(\bm{\lambda})\bigr)\Bigr),
  \label{cnuniv}\\
  &c^{\Xi}_{\cD}(\bm{\lambda})
  =\prod_{j=1}^M\frac{\tilde{c}_{d_j}(\bm{\lambda})}
  {\tilde{c}_{j-1}(\bm{\lambda})}\cdot\!\!
  \prod_{1\leq j<k\leq M}\frac{\beta_{j-1+k-1}(\bm{\lambda})}
  {\beta_{d_j+d_k}(\bm{\lambda})},
  \label{cXiD}\\
  &c^P_{\cD,n}(\bm{\lambda})
  =c^{\Xi}_{\cD}(\bm{\lambda})c_n(\bm{\lambda})
  \prod_{j=1}^M\frac{\beta'_{j-1}(\bm{\lambda})}{\beta'_{d_j+n}(\bm{\lambda})}.
  \label{cPDn}
\end{align}
We have $\mathfrak{t}^2=\text{id}$ and
$\mathfrak{t}(\bm{\lambda})+u\bm{\delta}
=\mathfrak{t}(\bm{\lambda}+u\tilde{\bm{\delta}})$ ($\forall u\in\mathbb{R}$).
For $\varphi(x)=q^{\pm x}$, we have
$\varphi_M(x)=q^{\pm(\binom{M}{2}x+\binom{M}{3})}$.
For a$q$K and $q$M cases, $\check{\xi}_{\text{v}}(x)$ \eqref{xivdef},
$B'(x)$ and $D'(x)$ \eqref{B'D'def} and $\cE'_{\text{v}}$ \eqref{E'vdef} are
defined without using the twist operation $\mathfrak{t}$.

\subsection{Each polynomial data}
\label{sec:eachpoly}

We assume that $\cD$ is standard order \eqref{stD}.
The range of parameters is expressed as (condition of the original system)
\& (condition of the deformed system).
The range of parameters may be extended.
The normalization constant $d_n^{\,2}$ is expressed as
$d_n^{\,2}=\frac{d_n^{\,2}}{d_0^{\,2}}\times d_0^{\,2}$.
Our normalizations of $\check{P}_n(x)$, $\phi_0(x)$, $\tilde{\phi}_0(x)$,
$\varphi(x)$, $\eta(x)$ and $\cE_n$ are
$\check{P}_n(0)=\phi_0(0)=\tilde{\phi}_0(0)=\varphi(0)=1$ and
$\eta(0)=\cE_0=0$.
The defining ranges of $(a)_x$ and $(a;q)_x$ can be extended to $x\in\mathbb{R}$
by $(a)_x=\Gamma(a+x)/\Gamma(a)$ and $(a;q)_x=(a;q)_{\infty}/(aq^x;q)_{\infty}$.

We have $\tcE_{\text{v}}<0$ (namely, $\tcE_{\text{v}}<\cE_n$
($\text{v}\in\cD$, $n\in\cX$)) except for dH, q$q$K and d$q$H types, for which
we have $\tcE_{\text{v}}>\cE_n$ ($\text{v}\in\cD$, $n\in\cX$).
The positivity of $B'(x)$, $D'(x)$, $\alpha$ and $-\alpha'$ is not
necessarily required.

Most of the data can be obtained from the data for $q$R case by taking
appropriate limits.

\subsubsection{Hahn (H)}
\label{sec:H}

The standard parametrization of Hahn polynomial \cite{kls} is
\begin{equation}
  (\alpha,\beta)^{\text{standard}}=(a-1,b-1).
\end{equation}

Basic data of the case-(1) multi-indexed Hahn polynomials are as follows:
\begin{align}
  &\bm{\lambda\,}\eqdef(a,b,N),\quad \bm{\delta}\eqdef(1,1,-1),
  \quad\kappa\eqdef 1,
  \label{lambda:H}\\
  &a>0,\ \ b>0,\ \ \&\ \ b>1+d_M,
  \label{range:H}\\
  &B(x;\bm{\lambda})\eqdef (x+a)(N-x),\quad
  D(x;\bm{\lambda})\eqdef x(b+N-x),
  \label{B,Ddef:H}\\
  &\cE_n(\bm{\lambda})\eqdef n(n+a+b-1),\quad
  \eta(x)\eqdef x,\quad
  \varphi(x)=1,
  \label{En,eta,varphidef:H}\\
  &\check{P}_n(x;\bm{\lambda})\eqdef
  {}_3F_2\Bigl(\genfrac{}{}{0pt}{}{-n,\,n+a+b-1,\,-x}{a,\,-N}\Bigm|1\Bigr)
  =Q_n\bigl(\eta(x);a-1,b-1,N\bigr),
  \label{Pndef:H}\\
  &\phi_0(x;\bm{\lambda})^2=
  \frac{(N-x+1)_x}{(1)_x}\frac{(a)_x}{(b+N-x)_x},
  \label{phi0:H}\\
  &d_n(\bm{\lambda})^2\eqdef
  \frac{(N-n+1)_n}{(1)_n}\frac{(a,a+b-1)_n}{(b,a+b+N)_n}
  \frac{2n+a+b-1}{a+b-1}\times
  \frac{(b)_N}{(a+b)_N},
  \ \ d_n(\bm{\lambda})>0,
  \label{dndef:H}\\
  &\mathfrak{t}(\bm{\lambda})\eqdef(a,2-b,N+b-1),\quad
  \tilde{\bm{\delta}}\eqdef(1,-1,0),
  \label{twistdef:H}\\
  &\alpha\eqdef 1,\quad
  \alpha'(\bm{\lambda})\eqdef-a(b-1),
  \label{alphadef:H}\\
  &\check{\xi}_{\text{v}}(x;\bm{\lambda})=
  {}_3F_2\Bigl(\genfrac{}{}{0pt}{}{-\text{v},\,\text{v}+a-b+1,\,-x}{a,1-N-b}
  \Bigm|1\Bigr),
  \label{xi:H}\\
  &\tcE_{\text{v}}(\bm{\lambda})=-(a+\text{v})(b-1-\text{v}),
  \label{Etv:H}\\
  &\nu(x;\bm{\lambda})=\frac{(N-x+1)_x}{(b+N-x)_x},
  \label{nu:H}\\
  &r_j(x_j;\bm{\lambda},M)=\frac{(N-x-j+2)_{j-1}(b+N-M-x)_{M+1-j}}{(b+N-M)_M},
  \label{rj:H}\\
  &c_n(\bm{\lambda})=\frac{(a+b+n-1)_n}{(a,-N)_n},\quad
  \tilde{c}_{\text{v}}(\bm{\lambda})
  =\frac{(a-b+\text{v}+1)_{\text{v}}}{(a,1-b-N)_{\text{v}}},
  \label{cn:H}\\
  &\beta_n(\bm{\lambda})\eqdef a-b+n+1,\quad
  \beta'_n(\bm{\lambda})\eqdef a+n,
  \label{betan:H}
\end{align}
where $Q_n$ in \eqref{Pndef:H} is the standard Hahn polynomial \cite{kls}.


\subsubsection{Racah (R)}
\label{sec:R}

The standard parametrization of Racah polynomial \cite{kls} is
\begin{equation}
  (\alpha,\beta,\gamma,\delta)^{\text{standard}}
  =(a-1,b+c-d-1,c-1,d-c).
\end{equation}
Basic data of the case-(1) multi-indexed Racah polynomials are as follows
\cite{os26}:
\begin{align}
  &\bm{\lambda\,}\eqdef(a,b,c,d),\quad \bm{\delta}\eqdef(1,1,1,1),
  \quad\kappa\eqdef 1,\quad
  \tilde{d}\eqdef a+b+c-d-1,
  \label{lambda:R}\\
  &a=-N,\ \ 0<d<a+b,\ \ 0<c<1+d,\ \ \&\ \ a+b>d+1+d_M,
  \label{range:R}\\
  &B(x;\bm{\lambda})\eqdef
  -\frac{(x+a)(x+b)(x+c)(x+d)}{(2x+d)(2x+1+d)},
  \label{Bdef:R}\\
  &D(x;\bm{\lambda})\eqdef
  -\frac{(x+d-a)(x+d-b)(x+d-c)x}{(2x-1+d)(2x+d)},
  \label{Ddef:R}\\
  &\cE_n(\bm{\lambda})\eqdef n(n+\tilde{d}),\quad
  \eta(x;\bm{\lambda})\eqdef x(x+d),\quad
  \varphi(x;\bm{\lambda})=\frac{2x+d+1}{d+1},
  \label{En,eta,varphidef:R}\\
  &\check{P}_n(x;\bm{\lambda})\eqdef
  {}_4F_3\Bigl(\genfrac{}{}{0pt}{}{-n,\,n+\tilde{d},\,-x,\,x+d}
  {a,\,b,\,c}\Bigm|1\Bigr)\n
  &\phantom{\check{P}_n(x;\bm{\lambda})}
  =R_n\bigl(\eta(x;\bm{\lambda});a-1,\tilde{d}-a,c-1,d-c\bigr),
  \label{Pndef:R}\\
  &\phi_0(x;\bm{\lambda})^2=
  \frac{(a,b,c,d)_x}{(1+d-a,1+d-b,1+d-c,1)_x}\,\frac{2x+d}{d},
  \label{phi0:R}\\
  &d_n(\bm{\lambda})^2\eqdef
  \frac{(a,b,c,\tilde{d})_n}{(1+\tilde{d}-a,1+\tilde{d}-b,1+\tilde{d}-c,1)_n}\,
  \frac{2n+\tilde{d}}{\tilde{d}}\n
  &\phantom{d_n(\bm{\lambda})^2\eqdef}
  \times\frac{(-1)^N(1+d-a,1+d-b,1+d-c)_N}{(\tilde{d}+1)_N(d+1)_{2N}},
  \ \ d_n(\bm{\lambda})>0,
  \label{dndef:R}\\
  &\mathfrak{t}(\bm{\lambda})\eqdef(d-a+1,d-b+1,c,d),\quad
  \tilde{\bm{\delta}}\eqdef(0,0,1,1),
  \label{twistdef:R}\\
  &\alpha\eqdef 1,\quad
  \alpha'(\bm{\lambda})\eqdef-c(a+b-d-1),
  \label{alphadef:R}\\
  &\check{\xi}_{\text{v}}(x;\bm{\lambda})=
  {}_4F_3\Bigl(\genfrac{}{}{0pt}{}{-\text{v},\,\text{v}-a-b+c+d+1,\,-x,\,x+d}
  {d-a+1,\,d-b+1,\,c}\Bigm|1\Bigr),
  \label{xi:R}\\
  &\tcE_{\text{v}}(\bm{\lambda})=-(c+\text{v})(a+b-d-1-\text{v}),
  \label{Etv:R}\\
  &\nu(x;\bm{\lambda})=\frac{(a,b)_x}{(d-a+1,d-b+1)_x},
  \label{nu:R}\\
  &r_j(x_j;\bm{\lambda},M)
  =\frac{(x+a,x+b)_{j-1}(x+d-a+j,x+d-b+j)_{M+1-j}}{(d-a+1,d-b+1)_M},
  \label{rj:R}\\
  &c_n(\bm{\lambda})=\frac{(\tilde{d}+n)_n}{(a,b,c)_n},\quad
  \tilde{c}_{\text{v}}(\bm{\lambda})
  =\frac{(c+d-a-b+\text{v}+1)_{\text{v}}}{(d-a+1,d-b+1,c)_{\text{v}}},
  \label{cn:R}\\
  &\beta_n(\bm{\lambda})\eqdef c+d-a-b+n+1,\quad
  \beta'_n(\bm{\lambda})\eqdef c+n,
  \label{betan:R}
\end{align}
where $R_n$ in \eqref{Pndef:R} is the standard Racah polynomial \cite{kls}.

\subsubsection{dual Hahn (dH)}
\label{sec:dH}

The standard parametrization of dual Hahn polynomial \cite{kls} is
\begin{equation}
  (\gamma,\delta)^{\text{standard}}=(a-1,b-1).
\end{equation}
Basic data of the case-(1) multi-indexed dual Hahn polynomials are as follows:
\begin{align}
  &\bm{\lambda\,}\eqdef(a,b,N),\quad \bm{\delta}\eqdef(1,0,-1),
  \quad\kappa\eqdef 1,
  \label{lambda:dH}\\
  &a>0,\ \ b>0,
  \label{range:dH}\\
  &B(x;\bm{\lambda})\eqdef
  \frac{(x+a)(x+a+b-1)(N-x)}{(2x-1+a+b)(2x+a+b)},
  \label{Bdef:dH}\\
  &D(x;\bm{\lambda})\eqdef\frac{x(x+b-1)(x+a+b+N-1)}{(2x-2+a+b)(2x-1+a+b)},
  \label{Ddef:dH}\\
  &\cE_n(\bm{\lambda})\eqdef n,\quad
  \eta(x;\bm{\lambda})\eqdef x(x+a+b-1),\quad
  \varphi(x;\bm{\lambda})=\frac{2x+a+b}{a+b},
  \label{En,eta,varphidef:dH}\\
  &\check{P}_n(x;\bm{\lambda})\eqdef
  {}_3F_2\Bigl(\genfrac{}{}{0pt}{}{-n,\,x+a+b-1,\,-x}{a,\,-N}\Bigm|1\Bigr)
  =R_n\bigl(\eta(x;\bm{\lambda});a-1,b-1,N\bigr),
  \label{Pndef:dH}\\
  &\phi_0(x;\bm{\lambda})^2=
  \frac{(N-x+1)_x}{(1)_x}\frac{(a,a+b-1)_x}{(b,a+b+N)_x}
  \frac{2x+a+b-1}{a+b-1},
  \label{phi0:dH}\\
  &d_n(\bm{\lambda})^2\eqdef
  \frac{(N-n+1)_n}{(1)_n}\frac{(a)_n}{(b+N-n)_n}\times
  \frac{(b)_N}{(a+b)_N},
  \ \ d_n(\bm{\lambda})>0,
  \label{dndef:dH}\\
  &\mathfrak{t}(\bm{\lambda})\eqdef(b,a,-a-b-N),\quad
  \tilde{\bm{\delta}}\eqdef(0,1,0),
  \label{twistdef:dH}\\
  &\alpha\eqdef 1,\quad
  \alpha'(\bm{\lambda})\eqdef b+N,
  \label{alphadef:dH}\\
  &\check{\xi}_{\text{v}}(x;\bm{\lambda})=
  {}_3F_2\Bigl(\genfrac{}{}{0pt}{}{-\text{v},\,x+a+b-1,\,-x}
  {b,\,a+b+N}\Bigm|1\Bigr),
  \label{xi:dH}\\
  &\tcE_{\text{v}}(\bm{\lambda})=b+N+\text{v},
  \label{Etv:dH}\\
  &\nu(x;\bm{\lambda})=\frac{(a,-N)_x}{(b,a+b+N)_x},
  \label{nu:dH}\\
  &r_j(x_j;\bm{\lambda},M)
  =\frac{(x-N,x+a)_{j-1}(a+b+N+x+j-1,b+x+j-1)_{M+1-j}}{(a+b+N,b)_M},
  \label{rj:dH}\\
  &c_n(\bm{\lambda})=\frac{1}{(a,-N)_n},\quad
  \tilde{c}_{\text{v}}(\bm{\lambda})
  =\frac{1}{(b,a+b+N)_{\text{v}}},
  \label{cn:dH}\\
  &\beta_n(\bm{\lambda})\eqdef 1,\quad
  \beta'_n(\bm{\lambda})\eqdef 1,
  \label{betan:dH}
\end{align}
where $R_n$ in \eqref{Pndef:dH} is the standard dual Hahn polynomial \cite{kls}.

\subsubsection{dual quantum $\bm{q}$-Krawtchouk (dq$\bm{q}$K)}
\label{sec:dqqK}

The dual quantum $q$-Krawtchouk polynomial is not treated in \cite{kls}.

Basic data of the case-(1) multi-indexed dual quantum $q$-Krawtchouk
polynomials are as follows:
\begin{align}
  &q^{\bm{\lambda}}\eqdef(p,q^N),\quad \bm{\delta}\eqdef(0,-1),
  \quad\kappa\eqdef q^{-1},
  \label{lambda:dqqK}\\
  &p>q^{-N},\ \ \&\ \ p>q^{-N-1-d_M},
  \label{range:dqqK}\\
  &B(x;\bm{\lambda})\eqdef p^{-1}q^{-x-N-1}(1-q^{N-x}),\quad
  D(x;\bm{\lambda})\eqdef(q^{-x}-1)(1-p^{-1}q^{-x}),
  \label{B,Ddef:dqqK}\\
  &\cE_n(\bm{\lambda})\eqdef q^{-n}-1,\quad
  \eta(x)\eqdef 1-q^x,\quad
  \varphi(x)=q^x,
  \label{En,eta,varphidef:dqqK}\\
  &\check{P}_n(x;\bm{\lambda})\eqdef
  {}_2\phi_1\Bigl(\genfrac{}{}{0pt}{}{q^{-n},\,q^{-x}}
  {q^{-N}}\Bigm|q\,;pq^{x+1}\Bigr),
  \label{Pndef:dqqK}\\
  &\phi_0(x;\bm{\lambda})^2=
  \frac{(q^{N-x+1};q)_x}{(q;q)_x}
  \frac{p^{-x}q^{-Nx}}{(p^{-1}q^{-x};q)_x},
  \label{phi0:dqqK}\\
  &d_n(\bm{\lambda})^2\eqdef
  \frac{(q^{N-n+1};q)_n}{(q;q)_n}
  \frac{p^{-n}q^{n(n-1-N)}}{(p^{-1}q^{-N};q)_n}
  \times(p^{-1}q^{-N};q)_N,
  \ \ d_n(\bm{\lambda})>0,
  \label{dndef:dqqK}\\
  &q^{\mathfrak{t}(\bm{\lambda})}\eqdef(q^{-N-1},p^{-1}q^{-1}),\quad
  \tilde{\bm{\delta}}\eqdef(1,0),
  \label{twistdef:dqqK}\\
  &\alpha(\bm{\lambda})\eqdef p^{-1}q^{-N-1},\quad
  \alpha'(\bm{\lambda})\eqdef-(1-p^{-1}q^{-N-1}),
  \label{alphadef:dqqK}\\
  &\check{\xi}_{\text{v}}(x;\bm{\lambda})=
  {}_2\phi_1\Bigl(
  \genfrac{}{}{0pt}{}{q^{-\text{v}},\,q^{-x}}{pq}\Bigm|q\,;q^{x-N}\Bigr),
  \label{xi:dqqK}\\
  &\tcE_{\text{v}}(\bm{\lambda})=-(1-p^{-1}q^{-N-1-\text{v}}),
  \label{Etv:dqqK}\\
  &\nu(x;\bm{\lambda})=\frac{(q^{-N};q)_x}{(pq;q)_x},\quad
  r_j(x_j;\bm{\lambda},M)
  =\frac{(q^{x-N};q)_{j-1}(pq^{x+j};q)_{M+1-j}}{(pq;q)_M},
  \label{nu,rj:dqqK}\\
  &c_n(\bm{\lambda})=\frac{p^nq^{-\binom{n}{2}}}{(q^{-N};q)_n},\quad
  \tilde{c}_{\text{v}}(\bm{\lambda})
  =\frac{q^{-N\text{v}-\frac12\text{v}(\text{v}+1)}}{(pq;q)_{\text{v}}},
  \label{cn:dqqK}\\
  &\beta_n(\bm{\lambda})\eqdef q^{-n},\quad
  \beta'_n(\bm{\lambda})\eqdef q^{-n}.
  \label{betan:dqqK}
\end{align}

\subsubsection{$\bm{q}$-Hahn ($\bm{q}$H)}
\label{sec:qH}

The standard parametrization of $q$-Hahn polynomial \cite{kls} is
\begin{equation}
  (\alpha,\beta)^{\text{standard}}=(aq^{-1},bq^{-1}).
\end{equation}
Basic data of the case-(1) multi-indexed $q$-Hahn polynomials are as follows:
\begin{align}
  &q^{\bm{\lambda}}\eqdef(a,b,q^N),\quad \bm{\delta}\eqdef(1,1,-1),
  \quad\kappa\eqdef q^{-1},
  \label{lambda:qH}\\
  &0<a<1,\ \ 0<b<1,\ \ \&\ \ b<q^{1+d_M},
  \label{range:qH}\\
  &B(x;\bm{\lambda})\eqdef(1-aq^x)(q^{x-N}-1),\quad
  D(x;\bm{\lambda})\eqdef aq^{-1}(1-q^x)(q^{x-N}-b),
  \label{B,Ddef:qH}\\
  &\cE_n(\bm{\lambda})\eqdef(q^{-n}-1)(1-abq^{n-1}),\quad
  \eta(x)\eqdef q^{-x}-1,\quad
  \varphi(x)=q^{-x},
  \label{En,eta,varphidef:qH}\\
  &\check{P}_n(x;\bm{\lambda})\eqdef
  {}_3\phi_2\Bigl(\genfrac{}{}{0pt}{}{q^{-n},\,abq^{n-1},\,q^{-x}}
  {a,\,q^{-N}}\Bigm|q\,;q\Bigr)
  =Q_n\bigl(1+\eta(x);aq^{-1},bq^{-1},N|q\bigr),
  \label{Pndef:qH}\\
  &\phi_0(x;\bm{\lambda})^2=
  \frac{(q^{N-x+1};q)_x}{(q;q)_x}
  \frac{(a;q)_x}{(bq^{N-x};q)_x\,a^x},
  \label{phi0:qH}\\
  &d_n(\bm{\lambda})^2\eqdef
  \frac{(q^{N-n+1};q)_n}{(q;q)_n}
  \frac{(a,abq^{-1};q)_n}{(b,abq^N;q)_n\,a^n}\,
  \frac{1-abq^{2n-1}}{1-abq^{-1}}
  \times\frac{(b;q)_N\,a^N}{(ab;q)_N},
  \ \ d_n(\bm{\lambda})>0,
  \label{dndef:qH}\\
  &q^{\mathfrak{t}(\bm{\lambda})}\eqdef(a,b^{-1}q^2,bq^{N-1}),\quad
  \tilde{\bm{\delta}}\eqdef(1,-1,0),
  \label{twistdef:qH}\\
  &\alpha(\bm{\lambda})\eqdef bq^{-1},\quad
  \alpha'(\bm{\lambda})\eqdef-(1-a)(1-bq^{-1}),
  \label{alphadef:qH}\\
  &\check{\xi}_{\text{v}}(x;\bm{\lambda})=
  {}_3\phi_2\Bigl(
  \genfrac{}{}{0pt}{}{q^{-\text{v}},ab^{-1}q^{\text{v}+1}\,q^{-x}}
  {a,\,b^{-1}q^{1-N}}\Bigm|q\,;q\Bigr),
  \label{xi:qH}\\
  &\tcE_{\text{v}}(\bm{\lambda})=-(1-aq^{\text{v}})(1-bq^{-\text{v}-1}),
  \label{Etv:qH}\\
  &\nu(x;\bm{\lambda})=
  \frac{(q^{N-x+1};q)_x}{(bq^{N-x};q)_x},\quad
  r_j(x_j;\bm{\lambda},M)=\frac{(q^{N-x-j+2};q)_{j-1}(bq^{N-M-x};q)_{M+1-j}}
  {(bq^{N-M};q)_M},
  \label{nu,rj:qH}\\
  &c_n(\bm{\lambda})=\frac{(abq^{n-1};q)_n}{(a,q^{-N};q)_n},,\quad
  \tilde{c}_{\text{v}}(\bm{\lambda})
  =\frac{(ab^{-1}q^{\text{v}+1};q)_{\text{v}}}{(a,b^{-1}q^{1-N};q)_{\text{v}}},
  \label{cn:qH}\\
  &\beta_n(\bm{\lambda})\eqdef 1-ab^{-1}q^{n+1},\quad
  \beta'_n(\bm{\lambda})\eqdef 1-aq^n,
  \label{betan:qH}
\end{align}
where $Q_n$ in \eqref{Pndef:qH} is the standard $q$-Hahn polynomial \cite{kls}.


\subsubsection{quantum $\bm{q}$-Krawtchouk (q$\bm{q}$K)}
\label{sec:qqK}

Basic data of the case-(1) multi-indexed quantum $q$-Krawtchouk
polynomials are as follows:
\begin{align}
  &q^{\bm{\lambda}}\eqdef(p,q^N),\quad \bm{\delta}\eqdef(1,-1),
  \quad\kappa\eqdef q,
  \label{lambda:qqK}\\
  &p>q^{-N},
  \label{range:qqK}\\
  &B(x;\bm{\lambda})\eqdef p^{-1}q^x(q^{x-N}-1),\quad
  D(x;\bm{\lambda})\eqdef(1-q^x)(1-p^{-1}q^{x-N-1}),
  \label{B,Ddef:qqK}\\
  &\cE_n\eqdef 1-q^n,\quad
  \eta(x)\eqdef q^{-x}-1,\quad
  \varphi(x)=q^{-x},
  \label{En,eta,varphidef:qqK}\\
  &\check{P}_n(x;\bm{\lambda})\eqdef
  {}_2\phi_1\Bigl(\genfrac{}{}{0pt}{}{q^{-n},\,q^{-x}}
  {q^{-N}}\Bigm|q\,;pq^{n+1}\Bigr)
  =K^{\text{qtm}}_n\bigl(1+\eta(x);p,N|q\bigr),
  \label{Pndef:qqK}\\
  &\phi_0(x;\bm{\lambda})^2=
  \frac{(q^{N-x+1};q)_x}{(q;q)_x}
  \frac{p^{-x}q^{x(x-1-N)}}{(p^{-1}q^{-N};q)_x},
  \label{phi0:qqK}\\
  &d_n(\bm{\lambda})^2\eqdef
  \frac{(q^{N-n+1};q)_n}{(q;q)_n}
  \frac{p^{-n}q^{-Nn}}{(p^{-1}q^{-n};q)_n}
  \times(p^{-1}q^{-N};q)_N,
  \ \ d_n(\bm{\lambda})>0,
  \label{dndef:qqK}\\
  &q^{\mathfrak{t}(\bm{\lambda})}\eqdef(p^{-1},pq^N),\quad
  \tilde{\bm{\delta}}\eqdef(-1,0),
  \label{twistdef:qqK}\\
  &\alpha(\bm{\lambda})\eqdef p^{-1},\quad
  \alpha'(\bm{\lambda})\eqdef 1-p^{-1},
  \label{alphadef:qqK}\\
  &\check{\xi}_{\text{v}}(x;\bm{\lambda})=
  {}_2\phi_1\Bigl(
  \genfrac{}{}{0pt}{}{q^{-\text{v}},\,q^{-x}}
  {p^{-1}q^{-N}}\Bigm|q\,;p^{-1}q^{\text{v}+1}\Bigr),
  \label{xi:qqK}\\
  &\tcE_{\text{v}}(\bm{\lambda})=1-p^{-1}q^{\text{v}},
  \label{Etv:qqK}\\
  &\nu(x;\bm{\lambda})=
  \frac{(q^{N+1-x};q)_x}{(pq^{N+1-x};q)_x},\quad
  r_j(x_j;\bm{\lambda},M)
  =\frac{(q^{N-x-j+2};q)_{j-1}(pq^{N-M-x+1};q)_{M+1-j}}{(pq^{N-M+1};q)_M},
  \label{nu,rj:qqK}\\
  &c_n(\bm{\lambda})=\frac{p^nq^{n^2}}{(q^{-N};q)_n},\quad
  \tilde{c}_{\text{v}}(\bm{\lambda})
  =\frac{p^{-\text{v}}q^{\text{v}^2}}{(p^{-1}q^{-N};q)_{\text{v}}},
  \label{cn:qqK}\\
  &\beta_n(\bm{\lambda})\eqdef q^n,\quad
  \beta'_n(\bm{\lambda})\eqdef q^n,
  \label{betan:qqK}
\end{align}
where $K^{\text{qtm}}_n$ in \eqref{Pndef:qqK} is the standard quantum
$q$-Krawtchouk polynomial \cite{kls}.

\subsubsection{affine $\bm{q}$-Krawtchouk (a$\bm{q}$K)}
\label{sec:aqK}

The quantities $\check{\xi}_{\text{v}}(x)$, $B'(x)$, $D'(x)$ and
$\cE'_{\text{v}}$ are defined without using the twist operation $\mathfrak{t}$.
Basic data of the case-(1) multi-indexed affine $q$-Krawtchouk
polynomials are as follows:
\begin{align}
  &q^{\bm{\lambda}}\eqdef(p,q^N),\quad \bm{\delta}\eqdef(1,-1),
  \quad\kappa\eqdef q^{-1},
  \label{lambda:aqK}\\
  &0<p<q^{-1},
  \label{range:aqK}\\
  &B(x;\bm{\lambda})\eqdef(q^{x-N}-1)(1-pq^{x+1}),\quad
  D(x;\bm{\lambda})\eqdef pq^{x-N}(1-q^x),
  \label{B,Ddef:aqK}\\
  &\cE_n\eqdef q^{-n}-1,\quad
  \eta(x)\eqdef q^{-x}-1,\quad
  \varphi(x)=q^{-x},
  \label{En,eta,varphidef:aqK}\\
  &\check{P}_n(x;\bm{\lambda})\eqdef
  {}_3\phi_2\Bigl(\genfrac{}{}{0pt}{}{q^{-n},\,q^{-x},\,0}
  {pq,\,q^{-N}}\Bigm|q\,;q\Bigr)
  =K^{\text{aff}}_n\bigl(1+\eta(x);p,N|q\bigr)\n
  &\phantom{\check{P}_n(x;\bm{\lambda})}
  =\frac{1}{(p^{-1}q^{-n};q)_n}\,
  {}_2\phi_1\Bigl(\genfrac{}{}{0pt}{}{q^{-n},\,q^{x-N}}
  {q^{-N}}\Bigm|q\,;p^{-1}q^{-x}\Bigr)
  \label{Pndef:aqK}\\
  &\phantom{\check{P}_n(x;\bm{\lambda})}
  =\frac{1}{(q^{N+1-n};q)_n}\,
  {}_2\phi_1\Bigl(\genfrac{}{}{0pt}{}{q^{-n},\,pq^{x+1}}
  {pq}\Bigm|q\,;q^{N+1-x}\Bigr),\n
  &\phi_0(x;\bm{\lambda})^2=
  \frac{(q^{N-x+1};q)_x}{(q;q)_x}
  \frac{(pq;q)_x}{(pq)^x},
  \label{phi0:aqK}\\
  &d_n(\bm{\lambda})^2\eqdef
  \frac{(q^{N-n+1};q)_n}{(q;q)_n}
  \frac{(pq;q)_n}{(pq)^n}
  \times(pq)^N,
  \ \ d_n(\bm{\lambda})>0,
  \label{dndef:aqK}\\
  &\tilde{\bm{\delta}}\eqdef(1,0),
  \label{twistdef:aqK}\\
  &B'(x;\bm{\lambda})\eqdef q^{x-N}(1-pq^{x+1}),\quad
  D'(x;\bm{\lambda})\eqdef pq(q^{x-N-1}-1)(1-q^x),
  \label{B'D'def:aqK}\\
  &\tilde{\phi}_0(x;\bm{\lambda})^2=
  \frac{(pq;q)_x}{(q^{N-x+1},q;q)_x(pq)^x},
  \label{phit0:aqK}\\
  &\alpha(\bm{\lambda})\eqdef 1,\quad
  \alpha'(\bm{\lambda})\eqdef-(1-pq),\quad
  \cE'_{\text{v}}(\bm{\lambda})\eqdef-pq(1-q^{\text{v}}),
  \label{alphaE'vdef:aqK}\\
  &\check{\xi}_{\text{v}}(x;\bm{\lambda})\eqdef
  {}_2\phi_1\Bigl(
  \genfrac{}{}{0pt}{}{q^{-\text{v}},\,q^{-x}}
  {pq}\Bigm|q\,;pq^{N+\text{v}+2}\Bigr),
  \label{xi:aqK}\\
  &\tcE_{\text{v}}(\bm{\lambda})=-(1-pq^{\text{v}+1}),
  \label{Etv:aqK}\\
  &\nu(x;\bm{\lambda})=(q^{N+1-x};q)_x,\quad
  r_j(x_j;\bm{\lambda},M)=(q^{N-x-j+2};q)_{j-1},
  \label{nu,rj:aqK}\\
  &c_n(\bm{\lambda})=\frac{1}{(pq,q^{-N};q)_n},\quad
  \tilde{c}_{\text{v}}(\bm{\lambda})
  =\frac{(pq^{N+\text{v}+1})^{\text{v}}}{(pq;q)_{\text{v}}},
  \label{cn:aqK}\\
  &\beta_n(\bm{\lambda})\eqdef q^n,\quad
  \beta'_n(\bm{\lambda})\eqdef 1-pq^{n+1},
  \label{betan:aqK}
\end{align}
where $K^{\text{aff}}_n$ in \eqref{Pndef:aqK} is the standard affine
$q$-Krawtchouk polynomial \cite{kls}.

\subsubsection{$\bm{q}$-Racah ($\bm{q}$R)}
\label{sec:qR}

The standard parametrization of $q$-Racah polynomial \cite{kls} is
\begin{equation}
  (\alpha,\beta,\gamma,\delta)^{\text{standard}}
  =(aq^{-1},bcd^{-1}q^{-1},cq^{-1},dc^{-1}).
\end{equation}
Basic data of the case-(1) multi-indexed $q$-Racah polynomials are as
follows \cite{os26}:
\begin{align}
  &q^{\bm{\lambda}}\eqdef(a,b,c,d),\quad \bm{\delta}\eqdef(1,1,1,1),
  \quad\kappa\eqdef q^{-1},\quad
  \tilde{d}\eqdef abcd^{-1}q^{-1},
  \label{lambda:qR}\\
  &a=q^{-N},\ \ 0<ab<d<1,\ \ qd<c<1,\ \ \&\ \ ab<dq^{1+d_M},
  \label{range:qR}\\
  &B(x;\bm{\lambda})\eqdef
  -\frac{(1-aq^x)(1-bq^x)(1-cq^x)(1-dq^x)}{(1-dq^{2x})(1-dq^{2x+1})},
  \label{Bdef:qR}\\
  &D(x;\bm{\lambda})\eqdef
  -\tilde{d}\,\frac{(1-a^{-1}dq^x)(1-b^{-1}dq^x)(1-c^{-1}dq^x)(1-q^x)}
  {(1-dq^{2x-1})(1-dq^{2x})},
  \label{Ddef:qR}\\
  &\cE_n(\bm{\lambda})\eqdef(q^{-n}-1)(1-\tilde{d}q^n),\n
  &\eta(x;\bm{\lambda})\eqdef(q^{-x}-1)(1-dq^x),\quad
  \varphi(x;\bm{\lambda})=\frac{q^{-x}-dq^{x+1}}{1-dq},
  \label{En,eta,varphidef:qR}\\
  &\check{P}_n(x;\bm{\lambda})\eqdef
  {}_4\phi_3\Bigl(\genfrac{}{}{0pt}{}{q^{-n},\,\tilde{d}q^n,\,q^{-x},\,dq^x}
  {a,\,b,\,c}\Bigm|q\,;q\Bigr)\n
  &\phantom{\check{P}_n(x;\bm{\lambda})}
  =R_n\bigl(1+d+\eta(x;\bm{\lambda});
  aq^{-1},\tilde{d}a^{-1},cq^{-1},dc^{-1}|q\bigr),
  \label{Pndef:qR}\\
  &\phi_0(x;\bm{\lambda})^2=
  \frac{(a,b,c,d\,;q)_x}{(a^{-1}dq,b^{-1}dq,c^{-1}dq,q\,;q)_x\,\tilde{d}^x}\,
  \frac{1-dq^{2x}}{1-d},
  \label{phi0:qR}\\
  &d_n(\bm{\lambda})^2\eqdef
  \frac{(a,b,c,\tilde{d}\,;q)_n}
  {(a^{-1}\tilde{d}q,b^{-1}\tilde{d}q,c^{-1}\tilde{d}q,q\,;q)_n\,d^n}\,
  \frac{1-\tilde{d}q^{2n}}{1-\tilde{d}}\n
  &\phantom{d_n(\bm{\lambda})^2\eqdef}
  \times
  \frac{(-1)^N(a^{-1}dq,b^{-1}dq,c^{-1}dq\,;q)_N\,\tilde{d}^Nq^{\frac12N(N+1)}}
  {(\tilde{d}q\,;q)_N(dq\,;q)_{2N}},
  \ \ d_n(\bm{\lambda})>0,
  \label{dndef:qR}\\
  &q^{\mathfrak{t}(\bm{\lambda})}\eqdef(a^{-1}dq,b^{-1}dq,c,d),\quad
  \tilde{\bm{\delta}}\eqdef(0,0,1,1),
  \label{twistdef:qR}\\
  &\alpha(\bm{\lambda})\eqdef abd^{-1}q^{-1},\quad
  \alpha'(\bm{\lambda})\eqdef-(1-c)(1-abd^{-1}q^{-1}),
  \label{alphadef:qR}\\
  &\check{\xi}_{\text{v}}(x;\bm{\lambda})=
  {}_4\phi_3\Bigl(
  \genfrac{}{}{0pt}{}{q^{-\text{v}},\,a^{-1}b^{-1}cdq^{\text{v}+1},
  \,q^{-x},\,dq^x}
  {a^{-1}dq,\,b^{-1}dq,\,c}\Bigm|q\,;q\Bigr),
  \label{xi:qR}\\
  &\tcE_{\text{v}}(\bm{\lambda})=-(1-cq^{\text{v}})(1-abd^{-1}q^{-1-\text{v}}),
  \label{Etv:qR}\\
  &\nu(x;\bm{\lambda})=(a^{-1}b^{-1}dq)^x
  \frac{(a,b;q)_x}{(a^{-1}dq,b^{-1}dq;q)_x},
  \label{nu:qR}\\
  &r_j(x_j;\bm{\lambda},M)=\frac{(aq^x,bq^x;q)_{j-1}
  (a^{-1}dq^{x+j},b^{-1}dq^{x+j};q)_{M+1-j}}
  {(abd^{-1}q^{-1})^{j-1}q^{Mx}(a^{-1}dq,b^{-1}dq;q)_M},
  \label{rj:qR}\\
  &c_n(\bm{\lambda})=\frac{(\tilde{d}q^n;q)_n}{(a,b,c;q)_n},\quad
  \tilde{c}_{\text{v}}(\bm{\lambda})
  =\frac{(a^{-1}b^{-1}cdq^{\text{v}+1};q)_{\text{v}}}
  {(a^{-1}dq,b^{-1}dq,c;q)_{\text{v}}},
  \label{cn:qR}\\
  &\beta_n(\bm{\lambda})\eqdef 1-a^{-1}b^{-1}cdq^{n+1},\quad
  \beta'_n(\bm{\lambda})\eqdef 1-cq^n,
  \label{betan:qR}
\end{align}
where $R_n$ in \eqref{Pndef:qR} is the standard $q$-Racah polynomial \cite{kls}.

\subsubsection{dual $\bm{q}$-Hahn (d$\bm{q}$H)}
\label{sec:dqH}

The standard parametrization of dual $q$-Hahn polynomial \cite{kls} is
\begin{equation}
  (\gamma,\delta)^{\text{standard}}=(aq^{-1},bq^{-1}).
\end{equation}
Basic data of the case-(1) multi-indexed dual $q$-Hahn polynomials are as
follows:
\begin{align}
  &q^{\bm{\lambda}}\eqdef(a,b,q^N),\quad \bm{\delta}\eqdef(1,0,-1),
  \quad\kappa\eqdef q^{-1},
  \label{lambda:dqH}\\
%
  &0<a<1,\ \ 0<b<1,
  \label{range:dqH}\\
  &B(x;\bm{\lambda})\eqdef
  \frac{(q^{x-N}-1)(1-aq^x)(1-abq^{x-1})}{(1-abq^{2x-1})(1-abq^{2x})},
  \label{Bdef:dqH}\\
  &D(x;\bm{\lambda})\eqdef
  aq^{x-N-1}\frac{(1-q^x)(1-abq^{x+N-1})(1-bq^{x-1})}
  {(1-abq^{2x-2})(1-abq^{2x-1})},
  \label{Ddef:dqH}\\
  &\cE_n(\bm{\lambda})\eqdef q^{-n}-1,
  \ \eta(x;\bm{\lambda})\eqdef(q^{-x}-1)(1-abq^{x-1}),
  \ \varphi(x;\bm{\lambda})=\frac{q^{-x}-abq^{x}}{1-ab},
  \label{En,eta,varphidef:dqH}\\
  &\check{P}_n(x;\bm{\lambda})\eqdef
  {}_3\phi_2\Bigl(\genfrac{}{}{0pt}{}{q^{-n},\,abq^{x-1},\,q^{-x}}
  {a,\,q^{-N}}\Bigm|q\,;q\Bigr)\n
  &\phantom{\check{P}_n(x;\bm{\lambda})}
  =R_n\bigl(1+abq^{-1}+\eta(x;\bm{\lambda});aq^{-1},bq^{-1},N|q\bigr),
  \label{Pndef:dqH}\\
  &\phi_0(x;\bm{\lambda})^2=
  \frac{(q^{N-x+1};q)_x}{(q;q)_x}
  \frac{(a,abq^{-1};q)_x}{(b,abq^N;q)_x\,a^x}
  \frac{1-abq^{2x-1}}{1-abq^{-1}},
  \label{phi0:dqH}\\
  &d_n(\bm{\lambda})^2\eqdef
  \frac{(q^{N-n+1};q)_n}{(q;q)_n}
  \frac{(a;q)_n}{(bq^{N-n};q)_n\,a^n}\,
  \times\frac{(b;q)_N\,a^N}{(ab;q)_N},
  \ \ d_n(\bm{\lambda})>0,
  \label{dndef:dqH}\\
  &q^{\mathfrak{t}(\bm{\lambda})}\eqdef(b,a,a^{-1}b^{-1}q^{-N}),\quad
  \tilde{\bm{\delta}}\eqdef(0,1,0),
  \label{twistdef:dqH}\\
  &\alpha(\bm{\lambda})\eqdef b^{-1}q^{-N},\quad
  \alpha'(\bm{\lambda})\eqdef b^{-1}q^{-N}-1,
  \label{alphadef:dqH}\\
  &\check{\xi}_{\text{v}}(x;\bm{\lambda})=
  {}_3\phi_2\Bigl(
  \genfrac{}{}{0pt}{}{q^{-\text{v}},\,abq^{x-1},\,q^{-x}}
  {b,\,abq^N}\Bigm|q\,;q\Bigr),
  \label{xi:dqH}\\
  &\tcE_{\text{v}}(\bm{\lambda})=b^{-1}q^{-N-\text{v}}-1,
  \label{Etv:dqH}\\
  &\nu(x;\bm{\lambda})
  =\frac{(q^{N+1-x},a;q)_x}{(abq^N,b^{-1}q^{1-x};q)_x},
  \label{nu:dqH}\\
  &r_j(x_j;\bm{\lambda},M)
  =\frac{(q^{x-N},aq^x;q)_{j-1}(abq^{N+x+j-1},bq^{x+j-1};q)_{M+1-j}}
  {(bq^N)^{1-j}q^{Mx}(abq^N,b;q)_M},
  \label{rj:dqH}\\
  &c_n(\bm{\lambda})=\frac{1}{(a,q^{-N};q)_n},\quad
  \tilde{c}_{\text{v}}(\bm{\lambda})=\frac{1}{(b,abq^N;q)_{\text{v}}},
  \label{cn:dqH}\\
  &\beta_n(\bm{\lambda})\eqdef 1,\quad
  \beta'_n(\bm{\lambda})\eqdef 1,
  \label{betan:dqH}
\end{align}
where $R_n$ in \eqref{Pndef:dqH} is the standard dual $q$-Hahn polynomial
\cite{kls}.

%

\subsubsection{Meixner (M)}
\label{sec:M}

Basic data of the case-(1) multi-indexed Meixner polynomials are as follows
\cite{os35}:
\begin{align}
  &\bm{\lambda}\eqdef(\beta,c),\quad \bm{\delta}\eqdef(1,0),
  \quad\kappa\eqdef1,
  \label{lambda:M}\\
  &\beta>0,\quad 0<c<1,
  \label{range:M}\\
  &B(x;\bm{\lambda})\eqdef c(x+\beta),\quad
  D(x)\eqdef x,
  \label{B,Ddef:M}\\
  &\cE_n(\bm{\lambda})\eqdef(1-c)n,\quad
  \eta(x)\eqdef x,\quad
  \varphi(x)=1,
  \label{En,eta,varphidef:M}\\
  &\check{P}_n(x;\bm{\lambda})\eqdef
  {}_2F_1\Bigl(\genfrac{}{}{0pt}{}{-n,\,-x}{\beta}\Bigm|1-c^{-1}\Bigr)
  =M_n(x;\beta,c),
  \label{Pndef:M}\\
  &\phi_0(x;\bm{\lambda})^2=\frac{(\beta)_x\,c^x}{(1)_x},
  \label{phi0:M}\\
  &d_n(\bm{\lambda})^2\eqdef
  \frac{(\beta)_n\,c^n}{(1)_n}\times(1-c)^{\beta},
  \ \ d_n(\bm{\lambda})>0,
  \label{dndef:M}\\
  &\mathfrak{t}(\bm{\lambda})\eqdef(\beta,c^{-1}),\quad
  \tilde{\bm{\delta}}\eqdef(1,0),
  \label{twistdef:M}\\
  &\alpha(\bm{\lambda})\eqdef c,\quad
  \alpha'(\bm{\lambda})\eqdef-(1-c)\beta,
  \label{alphadef:M}\\
  &\check{\xi}_{\text{v}}(x;\bm{\lambda})=
  {}_2F_1\Bigl(\genfrac{}{}{0pt}{}{-\text{v},\,-x}{\beta}\Bigm|1-c\Bigr),
  \label{xi:M}\\
  &\tcE_{\text{v}}(\bm{\lambda})=-(1-c)(\text{v}+\beta),
  \label{Etv:M}\\
  &\nu(x;\bm{\lambda})=c^x,\quad
  r_j(x_j;\bm{\lambda},M)=c^{j-1},
  \label{nu,rj:M}\\
  &c_n(\bm{\lambda})=\frac{(1-c^{-1})^n}{(\beta)_n},\quad
  \tilde{c}_{\text{v}}(\bm{\lambda})
  =\frac{(1-c)^{\text{v}}}{(\beta)_{\text{v}}},
  \label{cn:M}\\
  &\beta_n(\bm{\lambda})\eqdef 1,\quad
  \beta'_n(\bm{\lambda})\eqdef \beta+n,
  \label{betan:M}
\end{align}
where $M_n$ in \eqref{Pndef:M} is the standard Meixner polynomial \cite{kls}.


\subsubsection{little $\bm{q}$-Jacobi (l$\bm{q}$J)}
\label{sec:lqJ}

The standard parametrization of little $q$-Jacobi polynomial \cite{kls} is
\begin{equation}
  (a,b)^{\text{standard}}=(aq^{-1},bq^{-1}).
\end{equation}
Basic data of the case-(1) multi-indexed little $q$-Jacobi polynomials are
as follows \cite{os35}
(Note that the standard parametrization is used in \cite{os35}.):
\begin{align}
  &q^{\bm{\lambda}}\eqdef(a,b),\quad \bm{\delta}\eqdef(1,1),
  \quad\kappa\eqdef q^{-1},
  \label{lambda:lqJ}\\
  &0<a<1,\quad b<1,\ \ \&\ \ a<q^{1+d_M},
  \label{range:lqJ}\\
  &B(x;\bm{\lambda})\eqdef aq^{-1}(q^{-x}-b),\quad
  D(x)\eqdef q^{-x}-1,
  \label{B,Ddef:lqJ}\\
  &\cE_n(\bm{\lambda})\eqdef(q^{-n}-1)(1-abq^{n-1}),\quad
  \eta(x)\eqdef 1-q^x,\quad
  \varphi(x)=q^x,
  \label{En,eta,varphidef:lqJ}\\
  &\check{P}_n(x;\bm{\lambda})\eqdef
  {}_3\phi_1\Bigl(\genfrac{}{}{0pt}{}{q^{-n},\,abq^{n-1},\,q^{-x}}{b}\Bigm|
  q\,;a^{-1}q^{x+1}\Bigr)
  =c'_n(\bm{\lambda})\,p_n\bigl(1-\eta(x);aq^{-1},bq^{-1}|q\bigr)\n
  &\phantom{\check{P}_n(x;\bm{\lambda})}
  =c'_n(\bm{\lambda})\,{}_2\phi_1\Bigl(
  \genfrac{}{}{0pt}{}{q^{-n},\,abq^{n-1}}{a}\Bigm|q\,;q^{x+1}\Bigr),\quad
  c'_n(\bm{\lambda})\eqdef(-a)^{-n}q^{-\binom{n}{2}}\frac{(a;q)_n}{(b;q)_n},
  \label{Pndef:lqJ}\\
  &\phi_0(x;\bm{\lambda})^2=\frac{(b;q)_x}{(q;q)_x}a^x,
  \label{phi0:lqJ}\\
  &d_n(\bm{\lambda})^2\eqdef
  \frac{(b,ab;q)_n\,a^nq^{n(n-1)}}{(a,q;q)_n}\,\frac{1-abq^{2n-1}}{1-abq^{n-1}}
  \times\frac{(a;q)_{\infty}}{(ab;q)_{\infty}},
  \ \ d_n(\bm{\lambda})>0,
  \label{dndef:lqJ}\\
  &q^{\mathfrak{t}(\bm{\lambda})}\eqdef(a^{-1}q^2,b),\quad
  \tilde{\bm{\delta}}\eqdef(-1,1),
  \label{twistdef:lqJ}\\
  &\alpha(\bm{\lambda})\eqdef aq^{-1},\quad
  \alpha'(\bm{\lambda})\eqdef-(1-aq^{-1})(1-b),
  \label{alphadef:lqJ}\\
  &\check{\xi}_{\text{v}}(x;\bm{\lambda})=
  \frac{(aq^{-\text{v}-1};q)_{\text{v}}}{(b;q)_{\text{v}}}
  \,{}_2\phi_1\Bigl(
  \genfrac{}{}{0pt}{}{q^{-\text{v}},\,a^{-1}bq^{\text{v}+1}}{a^{-1}q^2}
  \Bigm|q\,;q^{x+1}\Bigr)\n
  &\phantom{\check{\xi}_{\text{v}}(x;\bm{\lambda})}
  =\frac{(aq^{-\text{v}-1};q)_{\text{v}}}{(b;q)_{\text{v}}}(bq^x;q)_{\text{v}}
  \,{}_3\phi_2\Bigl(
  \genfrac{}{}{0pt}{}{q^{-\text{v}},\,b^{-1}q^{1-\text{v}},\,0}
  {a^{-1}q^2,b^{-1}q^{1-\text{v}-x}}\Bigm|q\,;q\Bigr),
  \label{xi:lqJ}\\
  &\tcE_{\text{v}}(\bm{\lambda})=-(1-aq^{-\text{v}-1})(1-bq^{\text{v}}),
  \label{Etv:lqJ}\\
  &\nu(x;\bm{\lambda})=(aq^{-1})^x,\quad
  r_j(x_j;\bm{\lambda},M)=(aq^{-1})^{j-1}q^{Mx},
  \label{nu,rj:lqJ}\\
  &c_n(\bm{\lambda})=\frac{(-a)^{-n}q^{-n(n-1)}(abq^{n-1};q)_n}{(b;q)_n},\quad
  \tilde{c}_{\text{v}}(\bm{\lambda})
  =\frac{(-a)^{\text{v}}q^{-\text{v}(\text{v}+1)}
  (a^{-1}bq^{\text{v}+1};q)_{\text{v}}}{(b;q)_{\text{v}}},
  \label{cn:lqJ}\\
  &\beta_n(\bm{\lambda})\eqdef 1-ab^{-1}q^{-n-1},\quad
  \beta'_n(\bm{\lambda})\eqdef 1-b^{-1}q^{-n},
  \label{betan:lqJ}
\end{align}
where $p_n$ in \eqref{Pndef:lqJ} is the standard little $q$-Jacobi polynomial
\cite{kls}.

\subsubsection{little $\bm{q}$-Laguerre (l$\bm{q}$L)}
\label{sec:lqL}

The standard parametrization of little $q$-Laguerre polynomial \cite{kls} is
\begin{equation}
  a^{\text{standard}}=aq^{-1}.
\end{equation}
Basic data of the case-(1) multi-indexed little $q$-Laguerre polynomials are
as follows \cite{os35}
(Note that the standard parametrization is used in \cite{os35}.):
\begin{align}
  &q^{\bm{\lambda}}\eqdef a,\quad \bm{\delta}\eqdef 1,
  \quad\kappa\eqdef q^{-1},
  \label{lambda:lqL}\\
  &0<a<1,\ \ \&\ \ a<q^{1+d_M},
  \label{range:lqL}\\
  &B(x;\bm{\lambda})\eqdef aq^{-x-1},\quad
  D(x)\eqdef q^{-x}-1,
  \label{B,Ddef:lqL}\\
  &\cE_n(\bm{\lambda})\eqdef q^{-n}-1,\quad
  \eta(x)\eqdef 1-q^x,\quad
  \varphi(x)=q^x,
  \label{En,eta,varphidef:lqL}\\
  &\check{P}_n(x;\bm{\lambda})\eqdef
  {}_2\phi_0\Bigl(\genfrac{}{}{0pt}{}{q^{-n},\,q^{-x}}{-}\Bigm|
  q\,;a^{-1}q^{x+1}\Bigr)
  =c'_n(\bm{\lambda})\,p_n\bigl(1-\eta(x);aq^{-1}|q\bigr)\n
  &\phantom{\check{P}_n(x;\bm{\lambda})}
  =c'_n(\bm{\lambda})\,{}_2\phi_1\Bigl(
  \genfrac{}{}{0pt}{}{q^{-n},\,0}{a}\Bigm|q\,;q^{x+1}\Bigr),\quad
  c'_n(\bm{\lambda})\eqdef(-a)^{-n}q^{-\binom{n}{2}}(a;q)_n,
  \label{Pndef:lqL}\\
  &\phi_0(x;\bm{\lambda})^2=\frac{a^x}{(q;q)_x},
  \label{phi0:lqL}\\
  &d_n(\bm{\lambda})^2\eqdef\frac{a^nq^{n(n-1)}}{(a,q;q)_n}\,
  \times(a;q)_{\infty},\ \ d_n(\bm{\lambda})>0,
  \label{dndef:lqL}\\
  &q^{\mathfrak{t}(\bm{\lambda})}\eqdef a^{-1}q^2,\quad
  \tilde{\bm{\delta}}\eqdef -1,
  \label{twistdef:lqL}\\
  &\alpha(\bm{\lambda})\eqdef aq^{-1},\quad
  \alpha'(\bm{\lambda})\eqdef-(1-aq^{-1}),
  \label{alphadef:lqL}\\
  &\check{\xi}_{\text{v}}(x;\bm{\lambda})=
  (aq^{-\text{v}-1};q)_{\text{v}}
  \,{}_2\phi_1\Bigl(\genfrac{}{}{0pt}{}{q^{-\text{v}},\,0}{a^{-1}q^2}
  \Bigm|q\,;q^{x+1}\Bigr),
  \label{xi:lqL}\\
  &\tcE_{\text{v}}(\bm{\lambda})=-(1-aq^{-\text{v}-1}),
  \label{Etv:lqL}\\
  &\nu(x;\bm{\lambda})=(aq^{-1})^x,\quad
  r_j(x_j;\bm{\lambda},M)=(aq^{-1})^{j-1}q^{Mx},
  \label{nu,rj:lqL}\\
  &c_n(\bm{\lambda})=(-a)^{-n}q^{-n(n-1)},\quad
  \tilde{c}_{\text{v}}(\bm{\lambda})
  =(-a)^{\text{v}}q^{-\text{v}(\text{v}+1)},
  \label{cn:lqL}\\
  &\beta_n(\bm{\lambda})\eqdef q^{-n},\quad
  \beta'_n(\bm{\lambda})\eqdef q^{-n},
  \label{betan:lqL}
\end{align}
where $p_n$ in \eqref{Pndef:lqL} is the standard little $q$-Laguerre polynomial
\cite{kls}.

%

\subsubsection{$\bm{q}$-Meixner ($\bm{q}$M)}
\label{sec:qMI}

The quantities $\check{\xi}_{\text{v}}(x)$, $B'(x)$, $D'(x)$ and
$\cE'_{\text{v}}$ are defined without using the twist operation $\mathfrak{t}$.
Basic data of the case-(1) multi-indexed $q$-Meixner polynomials are as follows:
\begin{align}
  &q^{\bm{\lambda}}\eqdef(b,c),\quad \bm{\delta}\eqdef(1,-1),
  \quad\kappa\eqdef q,
  \label{lambda:qM}\\
  &0<b<q^{-1},\ \ c>0,
  \label{range:qM}\\
  &B(x;\bm{\lambda})\eqdef cq^x(1-bq^{x+1}),\quad
  D(x;\bm{\lambda})\eqdef(1-q^x)(1+bcq^x),
  \label{B,Ddef:qM}\\
  &\cE_n\eqdef 1-q^n,\quad
  \eta(x)\eqdef q^{-x}-1,\quad
  \varphi(x)=q^{-x},
  \label{En,eta,varphidef:qM}\\
  &\check{P}_n(x;\bm{\lambda})\eqdef
  {}_2\phi_1\Bigl(\genfrac{}{}{0pt}{}{q^{-n},\,q^{-x}}
  {bq}\Bigm|q\,;-c^{-1}q^{n+1}\Bigr)
  =M_n\bigl(1+\eta(x);b,c|q\bigr),
  \label{Pndef:qM}\\
  &\phi_0(x;\bm{\lambda})^2=
  \frac{(bq;q)_x}{(q,-bcq;q)_x}c^xq^{\binom{x}{2}},
  \label{phi0:qM}\\
  &d_n(\bm{\lambda})^2\eqdef
  \frac{q^n(bq;q)_n}{(q,-c^{-1}q;q)_n}
  \times\frac{(-bcq;q)_{\infty}}{(-c;q)_{\infty}},
  \ \ d_n(\bm{\lambda})>0,
  \label{dndef:qM}\\
  &\tilde{\bm{\delta}}\eqdef(1,0),
  \label{twistdef:qM}\\
  &B'(x;\bm{\lambda})\eqdef-(1-bq^{x+1})(1+bcq^{x+1}),\quad
  D'(x;\bm{\lambda})\eqdef-b^2cq^{x+1}(1-q^x),
  \label{B'D'def:qM}\\
  &\tilde{\phi}_0(x;\bm{\lambda})^2=
  \frac{(bq,-bcq;q)_x}{(q;q)_x}(b^2cq^2)^{-x}q^{-\binom{x}{2}},
  \label{phit0:qM}\\
  &\alpha(\bm{\lambda})\eqdef-b^{-1}q^{-1},\quad
  \alpha'(\bm{\lambda})\eqdef -(b^{-1}q^{-1}-1),\quad
  \cE'_{\text{v}}(\bm{\lambda})\eqdef q^{-\text{v}}-1,
  \label{alphaE'vdef:qM}\\
  &\check{\xi}_{\text{v}}(x;\bm{\lambda})\eqdef
  {}_3\phi_2\Bigl(
  \genfrac{}{}{0pt}{}{q^{-\text{v}},\,q^{-x},\,0}
  {bq,\,-bcq}\Bigm|q\,;q\Bigr),
  \label{xi:qM}\\
  &\tcE_{\text{v}}(\bm{\lambda})=-(b^{-1}q^{-\text{v}-1}-1),
  \label{Etv:qM}\\
  &\nu(x;\bm{\lambda})=\frac{1}{(-b^{-1}c^{-1}q^{-x};q)_x},\quad
  r_j(x_j;\bm{\lambda},M)=\frac{(-b^{-1}c^{-1}q^{-x-M};q)_{M+1-j}}
  {(-b^{-1}c^{-1}q^{-M};q)_M},
  \label{nu,rj:qM}\\
  &c_n(\bm{\lambda})=\frac{(-c)^{-n}q^{n^2}}{(bq;q)_n},\quad
  \tilde{c}_{\text{v}}(\bm{\lambda})
  =\frac{1}{(bq,-bcq;q)_{\text{v}}}
  \label{cn:qM}\\
  &\beta_n(\bm{\lambda})\eqdef 1,\quad
  \beta'_n(\bm{\lambda})\eqdef 1-bq^{n+1},
  \label{betan:qM}
\end{align}
where $M_n$ in \eqref{Pndef:qM} is the standard $q$-Meixner polynomial
\cite{kls}.




\end{document}